\documentclass[a4paper,11pt]{article}
\usepackage{jheppub}
\usepackage[T1]{fontenc}
\usepackage{enumerate}

\newcommand{\csch}{\mathrm{csch}}

\def\CF{\mathrm{C_F}}
\def\CFsq{\mathrm{C_F^2}}

\def\CA{\mathrm{C_A}}
\def\CAsq{\mathrm{C_A^2}}
\def\CAcub{\mathrm{C_A^3}}
\def\CAfour{\mathrm{C_A^4}}

\def\CAn{\mathrm{C_A}^{\!\!\!\!n}}
\def\CAnm{\mathrm{C_A}^{\!\!\!\!n-1}}

\def\W{\mathcal{W}}
\def\oW{\overline{\mathcal{W}}}
\def\tW{\widetilde{\mathcal{W}}}
\def\A{\mathcal{A}}
\def\B{\mathcal{B}}

\def\U{\mathcal{U}}

\def\cS{\mathcal{S}}
\def\inn{\mathrm{in}}
\def\out{\mathrm{out}}
\def\s{\sigma}
\def\S{\Sigma}
\def\X{{\scriptscriptstyle X}}
\def\R{{\scriptscriptstyle\mathrm{R}}}
\def\V{{\scriptscriptstyle\mathrm{V}}}

\newcommand{\mr}[1]{\mathrm{#1}}
\renewcommand{\d}{\mathrm{d}}
\newcommand{\as}{\alpha_s}
\newcommand{\Li}{\,\mathrm{Li}}
\def\Lb{\bar{L}}
\def\Lh{\widehat{L}}
\def\NG{\mathrm{NG}}
\def\irr{\mathrm{irr}}
\title{\boldmath Non-global logarithms at finite $N_c$ beyond\\ leading order}

\author[a]{Kamel Khelifa-Kerfa}
\author[b]{and Yazid Delenda}

\affiliation[a]{D\'{e}partement de Physique, Facult\'{e} des Sciences, Universit\'{e} Hassiba Benbouali de Chlef,\\
Chlef, Algeria}
\affiliation[b]{D\'{e}partement des Sciences de la Mati\`{e}re, Facult\'{e} des Sciences, Universit\'{e} Hadj Lakhdar,\\
Batna, Algeria}

\emailAdd{kamel.kkhelifa@gmail.com}
\emailAdd{yazid.delenda@gmail.com}

\abstract{We analytically compute non-global logarithms at \emph{finite} $N_c$ fully up to 4 loops and partially at 5 loops, for the hemisphere mass distribution in $e^+e^-\to$ di-jets to leading logarithmic accuracy. Our method of calculation relies solely on integrating the eikonal squared-amplitudes for the emission of soft energy-ordered real-virtual gluons over the appropriate phase space. We show that the series of non-global logarithms in the said distribution exhibits a pattern of \emph{exponentiation} thus confirming --- by means of brute force --- previous findings. In the large-$N_c$ limit, our results coincide with those recently reported in literature. A  comparison of our proposed exponential form with all-orders numerical solutions is performed and the phenomenological impact of the finite-$N_c$ corrections is discussed.}


\arxivnumber{1501.00475}

\begin{document}
\maketitle
\flushbottom

\section{Introduction}
\label{sec:intro}

As the LHC started colliding protons with unprecedented beam energy, interest has risen in topics that had not received much attention previously, with the aim of uncovering new physics signals. Of these topics, that have recently seen substantial development, is the substructure of ``fat'' jets originating from the almost-collinear decay products of heavy resonances that are highly boosted (see for example refs. \cite{Butterworth:2008iy, Kaplan:2008ie, Thaler:2008ju, Ellis:2009me,Gallicchio:2010sw, Gallicchio:2010dq, Thaler:2010tr, Cui:2010km, Abdesselam:2010pt, Gallicchio:2011xq, Altheimer:2012mn, Ellis:2012sn, Altheimer:2013yza}). Many substructure techniques, such as filtering \cite{Butterworth:2008iy}, pruning \cite{Ellis:2009me} and trimming \cite{Krohn:2009th}, have been developed for the purpose of improving the discrimination of signals from QCD background. The latter substructure techniques aid in providing cleaner and more accurate measurements of the properties of these resonances through: first, identifying the origin of the jet (decayed massive particle --- signal --- or plain QCD radiation --- background---). Second, mitigating away the jet constituent particles that have most likely originated from initial-state radiation, underlying-event and pile-up.

These techniques require in many situations calculations of QCD observables (e.g., jet mass, jet shapes, etc) which need special attention particularly in the vicinity of the threshold limit where they become highly contaminated with perturbative large logarithms as well as non-perturbative corrections.
There has recently been some analytical work on a handful of the said substructure techniques with the aim to pave the way to a better understanding of their analytical properties (see \cite{Dasgupta:2013ihk} and references therein). Nonetheless and for the majority of QCD observables and substructure techniques
the only other option available is resorting to numerical simulations which are based on Monte Carlo (MC) integration methods, and which use several approximations, e.g., \texttt{Herwig} \cite{Bahr:2008pv,Gieseke:2011na}, \texttt{Pythia} \cite{Sjostrand:2006za,Sjostrand:2007gs} and \texttt{Sherpa} \cite{Gleisberg:2008ta}. These MC event generators have been very successful in describing collider data and are commonly used in the extraction of crucial information to boost the search for new physics.

An important issue that needs addressing is the accuracy of the said MC algorithms and the range of validity of the approximations used therein. For instance, amongst the widely adopted approximations in the said MC generators is that of large-$N_c$ limit (with $N_c$ the dimension of $\mathrm{SU}(N_c)$ group). The latter limit, which corresponds to neglecting non-planar Feynman diagrams, greatly simplifies the otherwise tremendously complex colour structure, especially at high multiplicities. However, MC generators are generally tuned with data from collider experiments for parameters that account for non-perturbative effects such as hadronisation, underlying event, etc. The process of tuning is itself vulnerable to erroneously ascribing neglected perturbative (observable-dependent) components, which might be originating from finite-$N_c$ corrections, to universal non-perturbative parameters. This could then potentially be a source of major discrepancy between the data and the predictions by the MC generators. It is thus of great importance to assess the validity of these approximations and make sure that neglected terms would not affect precision measurements.

Amongst the issues that MC generators are meant to tackle is that of the resummation of large logarithms typically inherent in the distributions of most observables. These large logarithms are a manifestation of the miscancellation of infrared/collinear singularities at the matrix-element level, due to the exclusion of real-emission events in certain regions of phase space. For several observables of sufficiently inclusive nature,\footnote{By ``sufficiently inclusive'' one means observables that are inclusive over emissions in the entire angular phase space.} i.e., global observables, the resummation of these logarithms is relatively straightforward and has even been achieved analytically to NNNLL accuracy \cite{Hoang:2014wka}. In fact semi-numerical programs have been developed with the power of resumming a wide range of global observables up to NLL (\texttt{CAESAR} \cite{Banfi:2004yd}) and even to NNLL recently (\texttt{ARES} \cite{Banfi:2014sua}). However the extension of the resummation programme, up NNLL or even to just NLL accuracy in some cases, has seen slow progress for another class of observables, namely non-global observables \cite{Dasgupta:2001sh, Dasgupta:2002bw}.

Non-global observables are observables that are sensitive to emissions in restricted angular regions of the phase space. The distributions of such observables contain logarithms (named non-global logarithms (NGLs)) of the scales present in the process. For instance, the hemisphere mass distribution contains logarithms of the ratio $Q/(Q\rho)$\,, where $Q$ is the hard scale of the process and $\rho$ is the normalised hemisphere mass squared. In the region where $\rho\ll 1$, these NGLs can form large contributions to the said distributions, and should thus be resummed to all-orders. Up to very recently, their resummation was only  possible numerically at large $N_c$ by means of: an MC program \cite{Dasgupta:2001sh, Dasgupta:2002bw} or solutions to the non-linear integro-differential Banfi-Marchesini-Smye (BMS) evolution equation \cite{Banfi:2002hw}. The large-$N_c$ approximation significantly simplifies the colour flow in multiple gluon branchings enabling the possibility of the resummation of NGLs at least numerically. Much effort has recently been advocated to achieving numerical (analytical) resummation of NGLs at finite (large) $N_c$. The work of Hatta and Ueda \cite{Hatta:2013iba} exploits the suggestion of Weigert \cite{Weigert:2003mm} to use an analogy between the resummation of small-Bjorken-$x$ (BFKL) logarithms and that of NGLs at finite $N_c$ in a numerical fashion. They have noticed that the neglected finite-$N_c$ corrections are indeed negligible in the context of $e^+e^- \to$ di-jets. They have however speculated that the situation may be drastically different for hadronic collisions. Furthermore, Rubin \cite{Rubin:2010fc} numerically computed the NGLs series for both filtered Higgs-jet mass as well as interjet energy flow observables up to five- and six-loops, respectively, at large $N_c$. In the same limit, Schwartz and Zhu \cite{Schwartz:2014wha} worked on the analytical solution to the BMS equation by means of an iterative series-solution up to five-loops.

The major hindrance that one inevitably faces when attempting to compute NGLs analytically at finite $N_c$ is twofold. Firstly, the colour topology of a multi-gluon event requires evaluations of non-trivial traces of colour matrices in $\mathrm{SU}(N_c)$, which become increasingly cumbersome starting from four-loops. Secondly and not less important, the non-Abelian gluon branchings increase the number of Feynman diagrams factorially at each escalating order to the extent that an automated way of accounting for all possible branchings becomes inescapable.\footnote{The number of cut diagrams to consider at $n$ loops is formally $((n+1)!)^2$ for real gluon emission. This number is slightly reduced by considering on-shell particles and exploiting available symmetries.} Besides, there is also the issue of the various possible real, virtual and real-virtual gluon configurations that are eventually responsible for the miscancellation of soft singularities, thus leading to the appearance of large logarithms. These difficulties may have been the main reason for the slow progress in the resummation of NGLs at finite $N_c$.

In this paper we overcome the above-mentioned difficulties and present the first analytical calculation of NGLs at \emph{finite} $N_c$ beyond leading order. Working in the eikonal approximation \cite{Levy:1969cr, Dokshitzer:1991wu, Bassetto:1984ik, Catani:1999ss, Catani:1996vz, Catani:1996jh} for soft (strongly) energy-ordered partons, the first problem, i.e, that of colour structure, is resolved via the use of the \texttt{Mathematica} package \texttt{ColorMath} developed by Sj\"{o}dahl \cite{Sjodahl:2012nk,Sjodahl:2013hra}. The latter program performs the summation of $\mathrm{SU}(N_c)$ colour matrices in an automated way at any loop order. For the second obstacle we developed a {\tt Mathematica} code that accounts for all possible gluon branchings (and thus for all possible antenna functions) in an automated way.\footnote{This code will be improved, in the near future \cite{Delenda:eikamp}, into a full {\tt Mathematica} package capable of analytically computing QCD eikonal amplitudes at (theoretically) any loop order.} Consequently we have been able to analytically calculate all squared amplitudes for the emission of soft energy-ordered gluons (for all possible real, virtual and real-virtual configurations) in the eikonal approximation fully (at finite $N_c$) up to five-loops. We leave the presentation of these squared-amplitudes and the method of calculation to a forthcoming paper \cite{Delenda:eikamp}.

With these squared-amplitudes at hand, we provide in this paper a calculation of NGLs at finite $N_c$ to single logarithmic accuracy for single-hemisphere mass distribution in $e^+e^-\to$ di-jets up to five-loops. While our calculation is full at four-loops it is incomplete at five-loops due to missing terms for which the squared amplitudes are not so simple to simplify and/or integrate. We find that the aforementioned distribution exhibits a pattern of exponentiation both for global and non-global logarithms. We consequently write the all-orders distribution as a product of two exponentials; the first being the usual Sudakov form factor and the second represents the ``resummed form factor'' for NGLs. For the sake of cross-checking we take the large-$N_c$ limit of our result and compare it with previous calculations obtained by Schwartz and Zhu \cite{Schwartz:2014wha}. We find complete agreement up to our accuracy, which is four-loops. Furthermore, we compare our analytical resummed factor to the available all-orders numerical results \cite{Dasgupta:2001sh} and discuss the phenomenological implications of our findings, particularly the issue of the accuracy of the large-$N_c$ limit, by assessing the importance of neglected finite-$N_c$ corrections up to four-loops.

This paper is organised as follows. In the next section we outline the usual procedure of calculating NGLs by defining the observable, kinematics and the general relation for the hemisphere mass distribution in terms of the squared amplitudes and a ``measurement operator''.\footnote{The idea of the ``measurement operator'' was introduced by Schwartz and Zhu in their paper \cite{Schwartz:2014wha} which we found very helpful in organizing the real-virtual contributions to NGLs.} We present, in the same section, the calculation of NGLs at leading order (two-loops) to warm up for higher loops. In section 3 we explicitly calculate NGLs beyond leading order at three, four and five-loops. The difficulties associated with calculations at five-loops will be addressed therein. We compare our findings, in section 4, to those obtained at large $N_c$ in ref. \cite{Schwartz:2014wha} as well as to the all-orders parametrised form reported by Dasgupta and Salam, which they obtained by fitting to the output of their MC program \cite{Dasgupta:2001sh}. We also assess the relative size of the corrections due to finite $N_c$ up to four-loops and discuss our findings in the same section. Finally, we conclude our work in section 5.

\section{Hemisphere mass distribution at one and two-loops}
\label{sec:hemisphere}

Our aim in this paper is to calculate NGLs at finite $N_c$ to single logarithmic accuracy up to the fifth order in the strong coupling $\alpha_s$ (or equivalently up to five-loops). Our calculation is performed using QCD squared-amplitudes for the emission of energy-ordered gluons in the eikonal approximation. The latter is sufficient to capture all single logarithms $\alpha_s^n L^n$, with $L$ being the large NGL. As stated in the introduction, we do not show herein explicit formulae for the said squared-amplitudes and refer the reader to our coming paper \cite{Delenda:eikamp}. Moreover, for the purpose of this paper, we do not consider the role of any jet algorithm, and postpone such work to future publications.

\subsection{Observable definition and kinematics}

For the sake of illustration and to avoid unnecessary complications from a hadronic environment, we choose to work with the same observable that was used in the original paper on NGLs by Dasgupta and Salam \cite{Dasgupta:2001sh} within the framework of QCD, that is, the hemisphere mass distribution in $e^+e^-\to$ di-jets. This very observable was also considered in refs. \cite{Kelley:2011ng,Hornig:2011iu,Hoang:2008fs} in the context of soft-collinear effective theory (SCET). In both refs. \cite{Dasgupta:2001sh,Kelley:2011ng,Hornig:2011iu, Hoang:2008fs} NGLs were only computed up to two-loops. In the eikonal approximation, sufficient for our purpose, we consider energy-ordered soft gluon emissions:\footnote{Since gluons must satisfy Bose statistics one should normally allow for the permutations of the gluons and divide by a factor $n!$. This is however equivalent to choosing a specific ordering and removing the $1/n!$ factor.} $Q\gg k_{t1}\gg k_{t2} \gg \cdots \gg k_{tn}$\,, with $Q$ the centre of mass energy and $k_{ti}$ the transverse momenta of emitted gluons $k_i$. We note that gluon decay into quarks has a sub-leading contribution to NGLs as was found at two-loops in refs. \cite{Kelley:2011ng, KhelifaKerfa:2011zu}.

The four-momenta of the outgoing quark, anti-quark and gluons are expressed in rapidity parametrisation as:
\begin{subequations}
\begin{align}
p_{q} =& \frac{Q}{2}(1,0,0,1)\,, \\
p_{\bar{q}} =& \frac{Q}{2}(1,0,0,-1)\,, \\
k_{i} =& k_{ti}(\cosh\eta_i,\cos\phi_i,\sin\phi_i,\sinh\eta_i)\,,
\end{align}
\end{subequations}
where recoil effects are negligible to single logarithmic accuracy. Here $\eta_i$ and $\phi_i$ are the rapidity and azimuthal angle of the $i^{\mathrm{th}}$ emission and $k_{ti}$ its transverse momentum with respect to the $z$-axis, which we choose to coincide with the outgoing quark direction. We have $k_{ti}=\omega_i \sin\theta_i$\,, with $\omega_i$ the energy of gluon $k_i$ and $\theta_i$ its polar angle. The rapidity is related to the polar angle $\theta_i$ through the relation $\eta_i = - \ln \tan(\theta_i/2)$.

\begin{figure}[t]
\centering
\includegraphics{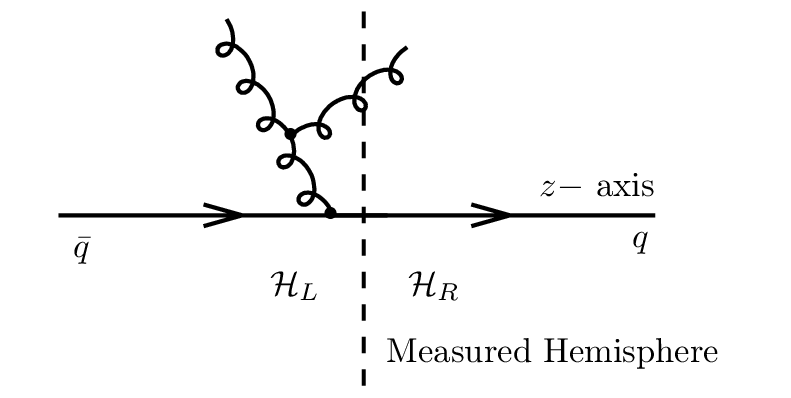}
\caption{\label{fig:setup}Schematic diagram for an outgoing $q\bar{q}$ pair associated with multiple gluon emission. The measured hemisphere is the one pointing in the quark direction ($\mathcal{H}_R$).}
\end{figure}

We define the following ``antenna'' functions relevant to the squared amplitudes that we use in this paper:
\begin{subequations}\label{eq:antennas}
\begin{align}
w_{ab}^i& = k_{ti}^2 \frac{p_a.p_b}{(p_a.k_i)\,(k_i.p_b)}\,,\\
\A_{ab}^{ij}& =w_{ab}^i \left(w_{ai}^j+w_{ib}^j-w_{ab}^j\right),\\
\B_{ab}^{ijk}& =w_{ab}^i \left(\A_{ai}^{jk}+\A_{ib}^{jk}-\A_{ab}^{jk}\right).
\end{align}
\end{subequations}
The quark and anti-quark directions define two coaxial back-to-back hemispheres ($\mathcal{H}_L$ and $\mathcal{H}_R$) whose axis coincides with the thrust axis at single logarithmic accuracy (see figure \ref{fig:setup}).
We pick for measurement the hemisphere pointing in the positive $z$-axis (quark direction). The normalised hemisphere mass (squared) $\rho$ is then defined by:
\begin{equation}
\begin{split}
\rho &= \left(p_q+\sum_{i\in\mathcal{H}_R} k_i \right)^2/Q^2  \approx\, 2 \sum_{i\in\mathcal{H}_R } k_i. p_q/Q^2 = \sum_i \rho_i\,,\\
 \rho_i &\equiv \,2\, k_i\cdot p_q/Q^2 = x_i \, e^{-\eta_i} \,,
\end{split}
\end{equation}
where we introduced the transverse momenta fractions $x_i = k_{ti}/Q$\,, and the sum over the index $i$ extends over all emitted \emph{real} gluons in the measured hemisphere $\mathcal{H}_R$.

We compute the integrated hemisphere mass distribution (cross-section) normalised to the Born cross-section, defined by:
\begin{align}
\S(\rho) &= \int_0^\rho \frac{1}{\s_0} \frac{\d\s}{\d\rho'}\,\d\rho'\notag \\
&= 1+\S_1(\rho) + \S_2(\rho)+\cdots\,, \label{eq:hemi_mass_fraction}
\end{align}
with
\begin{align}\label{eq:Sigma_m}
\S_m(\rho) = \sum_\X \int_{x_1>x_2>\cdots >x_m} \left(\prod_{i=1}^m \d\widetilde{\Phi}_i\right)  \hat{\U}_m \, \tW^{\X}_{12\cdots m}\,,
\end{align}
where $\tW^{\X}_{12\cdots m}=\tW^{\X}(k_1,k_2,\cdots, k_m)$ is the eikonal matrix-element squared for the emission of $m$ energy-ordered soft gluons of configuration $X$ off the primary $q\bar{q}$ pair at $m^{\mathrm{th}}$ order, normalised to the Born cross-section. The sum over $X$ extends over all possible real (R) and/or virtual (V) configurations of all the gluons $\{k_j\}$. For instance, at 2 loops ($m=2$) the eikonal squared-amplitudes $\tW^{\X}_{12}$ over which the sum is taken are: $\tW_{12}^{\R\R}$, $\tW_{12}^{\R\V}$, $\tW_{12}^{\V\R}$, and $\tW_{12}^{\V\V}$.
The quantity $\tW_{12}^{\R\V}$, for example, is read as: the squared amplitude for the emission of two energy-ordered gluons, $k_1$ and $k_2$, with gluon $k_1$ real and gluon $k_2$ virtual. Notice that, in the eikonal approximation, the squared amplitude for the softest gluon being virtual is simply minus the squared amplitude for it being real. In other words:
\begin{align}
\tW_{12 \cdots m}^{\mr{x x} \cdots \V} = - \tW^{\mr{x x} \cdots \R}_{12 \cdots m}\,,
\end{align}
where $\mr{x}$ could either be R or V. At one- and two-loops, for example, one has:
\begin{equation}
\begin{split}
\tW^{\V}_{1} = & - \tW^{\R}_{1}\,, \\
 \tW^{\R\V}_{12} = & - \tW^{\R\R}_{12}\,, \qquad   \tW^{\V\V}_{12} = - \tW^{\V\R}_{12}\,.
\end{split}
\end{equation}
The phase space factor for the emission of $m$ gluons is:
\begin{align}
\prod_{i=1}^m  \d\widetilde{\Phi}_i &= \prod_{i=1}^m \frac{\d^3 k_i}{(2\pi)^3 2 \omega_i}
= \bar{\alpha}_s^m \prod_{i=1}^m  \frac{\d x_i}{x_i} \d \eta_i \frac{\d\phi_i}{2\pi} \frac{k_{ti}^2}{2g_s^2}\,,
\end{align}
where $g_s=\sqrt{4\pi \alpha_s}$ and $\bar{\alpha}_s = \alpha_s/\pi$. The factor $\prod_{i=1}^m k_{ti}^2/2g_s^2$ multiplies the squared amplitude $\tW^{\X}_{12\cdots m}$ to produce the \emph{purely} angular squared-amplitude $\W^{\X}_{12\cdots m}$ (i.e., $\W^{\X}_{12\cdots m}$ depends only on $\eta$ and $\phi$ variables and the corresponding colour factor). In other words, one may write:
\begin{align}
&\prod_{i=1}^m  \d\widetilde{\Phi}_i\, \tW^{\X}_{12\cdots m} = \prod_{i=1}^m  \d\Phi_i\, \W^{\X}_{12\cdots m}\,,
\end{align}
where
\begin{equation}
\begin{split}
\prod_{i=1}^m\d\Phi_i &= \prod_{i=1}^m \d\widetilde{\Phi}_i  \frac{2g_s^2}{k_{ti}^2}=\bar{\alpha}_s^m \prod_{i=1}^m  \frac{\d x_i}{x_i} \d \eta_i \frac{\d\phi_i}{2\pi}\,,\\
\W^{\X}_{12\cdots m} &= \tW^{\X}_{12\cdots m} \prod_{i=1}^m\frac{k_{ti}^2}{2g_s^2}\,.
\end{split}
\end{equation}
The \emph{non-linear} ``measurement operator'' $\hat{\U}_m$ acts on the squared amplitudes $\W_{12\cdots m}^{\X}$ and plays the role of \emph{excluding} gluon emission events for which the hemisphere mass is greater than $\rho$. It is not, however, equivalent to a simple heaviside step function $\Theta(\rho - \sum_i x_i e^{-\eta_i})$, since it requires non-numerical input (information about the real-virtual nature of the various gluons). Due to strong ordering, the measurement operator $\hat{\U}_m$ factorises into a product of individual measurement operators; $\hat{\U}_m = \prod_{i=1}^m \hat{u}_i$.
The squared amplitudes $\W_{12\cdots m}^{\X}$ are eigenfunctions of the measurement operators $\hat{u}_i$ with eigenvalues $0$ or $1$ such that:
\begin{itemize}
\item if gluon $k_i$ is virtual then $\hat{u}_i\,\W_{12\cdots m}^{\X} = \W_{12\cdots m}^{\X}\,$,
\item if gluon $k_i$ is real and outside $\mathcal{H}_R$ then $\hat{u}_i\,\W_{12\cdots m}^{\X} = \W_{12\cdots m}^{\X}\,$,
\item if gluon $k_i$ is real and inside $\mathcal{H}_R$ with $\rho_i<\rho$ then $\hat{u}_i\,\W_{12\cdots m}^{\X} = \W_{12\cdots m}^{\X}\,$,
\item if gluon $k_i$ is real and inside $\mathcal{H}_R$ with $\rho_i>\rho$ then $\hat{u}_i \,\W_{12\cdots m}^{\X} = 0\,$.
\end{itemize}
This means that events with real emissions inside the measured hemisphere and which contribute more than $\rho$ to the hemisphere mass are excluded (i.e., \emph{not} integrated over). That is, $\S(\rho)$ represents the probability that the measured hemisphere mass be less than $\rho$\,, as is expressed in eq. \eqref{eq:hemi_mass_fraction}. We therefore write the measurement operator as:
\begin{align}
\hat{u}_i =& \hat{\Theta}^\V_i+\hat{\Theta}^\R_i\left[\Theta^\out_i
+\Theta^\inn_i\Theta\left(\rho - \rho_i\right)\right]= 1- \Theta^\rho_i \Theta^{\inn}_i\hat{\Theta}^\R_i\,,
\label{eq:meas_operator_ui}
\end{align}
with $\Theta^\rho_i = \Theta(\rho_i-\rho) = \Theta(x_i e^{-\eta_i}-\rho) $\,,  $\Theta^{\inn}_i = \Theta(\eta_i) $\,, and $\Theta^\out_i = \Theta(-\eta_i)$\,. The heaviside step functions $\Theta^\inn_i$ and $\Theta^\out_i$ respectively indicate whether gluon $k_i$ is inside or outside the measured hemisphere region. The operator $\hat{\Theta}^\R_i$ ($\hat{\Theta}^\V_i$) equals $1$ if gluon $k_i$ is real (virtual) and $0$ otherwise. If gluon $k_i$ is real then $\hat{\Theta}^\R_i \W_{12\cdots m}^{\X} = \W_{12\cdots m}^{\X}$ and $\hat{\Theta}^\V_i \W_{12\cdots m}^{\X} = 0$, and vice versa. In the above we used the relations $\Theta^{\inn}_i+\Theta^\out_i = 1$ and $\hat{\Theta}^\R_i + \hat{\Theta}^\V_i= 1$.

\subsection{One-loop calculation and the Sudakov exponentiation}

Having properly set up the form of the integrated distribution, we can now proceed with the calculation at each loop level. To warm up for higher loops we start with the one-loop case.
At one-loop the squared amplitude for the emission of a single real gluon, multiplied by the corresponding phase space, is given by:
\begin{align}\label{eq:M1}
\d\Phi_1 \times \W_1^{\R} =& \bar{\alpha}_s \frac{\d x_1}{x_1} \d \eta_1 \frac{\d\phi_1}{2\pi} \times \CF w_{q\bar{q}}^1\notag\\
=&  \bar{\alpha}_s \frac{\d x_1}{x_1} \d \eta_1 \frac{\d\phi_1}{2\pi} 2\,\CF\,.
\end{align}
The corresponding virtual contribution is $\W_1^{\V}=- \W_1^{\R}$. The measurement operator reads:
\begin{align} \label{eq:u(k1)}
\hat{\mathcal{U}}_1 = \hat{u}_1 &= 1- \Theta^\rho_1\Theta^{\inn}_1\hat{\Theta}_1^\R \,,
\end{align}
which when acting on the squared eikonal amplitudes yields:
\begin{align}
\hat{u}_1 \W_1^\R+ \hat{u}_1 \W_1^\V &=  \W_1^\R-\Theta^\rho_1 \Theta^{\inn}_1 \hat{\Theta}^\R_1 \W_1^\R +\W_1^\V - \Theta^\rho_1 \Theta^{\inn}_1 \hat{\Theta}^\R_1 \W_1^\V \notag\\
&= -\Theta^\rho_1 \Theta^{\inn}_1 \W_1^\R\,,
\end{align}
where we used $\hat{\Theta}^\R_1 \W_1^\V = 0$\,, $\hat{\Theta}^\R_1 \W_1^\R=\W_1^\R$ and $\W_1^\R+\W_1^\V =0$ (which means that the real and virtual contributions completely cancel out in sufficiently \emph{inclusive} cross-sections).
Substituting into the expression of $\S_1$ (e.q. \eqref{eq:Sigma_m} with $m=1$), we are left with the uncancelled integration:
\begin{align}
\S_1(\rho) &  =- \int \d\Phi_1 \Theta^\rho_1 \Theta^\inn_1 \W_1^{\R} \notag\\
 & = -2\,\CF\bar{\alpha}_s \int_0^{L} \d \eta_1 \int^1_{\rho \, e^{\eta_1}} \frac{\d x_1}{x_1}   \int_0^{2\pi}\frac{\d\phi_1}{2\pi}\,,\label{eq:S1}
\end{align}
where the step function $\Theta^\rho_1$ restricts $x_1$ to be greater than $\rho \, e^{\eta_1}$, and since $x_1<1$ then one also has the restriction on the rapidity such that $\eta_1<L$, with $L=\ln(1/\rho)$.
Performing the integration to single logarithmic accuracy we find:
\begin{align} \label{eq:S1_b}
\S_1 (\rho) & = -\CF\bar{\alpha}_s L^2 \equiv \S_1^\mathrm{P}(\rho)\,.
\end{align}
We note that the leading logarithms in the hemisphere mass distribution are double logarithms, which originate from soft and collinear (to the direction of the outgoing quark) singularities of the squared amplitudes for primary gluon emissions off the initiating hard $q\bar{q}$ pair.
It has long been known that the resummed distribution accounting for these primary emissions (or global logarithms) to all-orders is entirely generated from the leading-order result by simple exponentiation (Sudakov form factor). However, Dasgupta and Salam \cite{Dasgupta:2001sh} showed that a new class of large single logarithms appears starting at two gluons emission, and which they termed NGLs. We may thus express the resummed hemisphere mass distribution as follows:
\begin{equation}\label{eq:splitting}
\S(\rho) =\S^{\mathrm{P}}(\rho)\times  \S^{\NG}(\rho)\,,
\end{equation}
where $\S^{\mr{P}}$ is the primary Sudakov form factor,
\begin{align}
\S^{\mathrm{P}}(\rho) &= 1+\S_1^\mathrm{P}+\frac{1}{2!}\left(\S_1^{\mathrm{P}}\right)^2 +\frac{1}{3!}\left(\S_1^{\mathrm{P}}\right)^3 + \cdots \notag\\
&= \exp\left(\S_1^\mathrm{P}\right)=  \exp\left( -\CF \bar{\alpha}_sL^2 \right),
\end{align}
and $\S^{\NG}$ is the resummed non-global factor,
\begin{align}
\S^{\NG}(\rho) = 1+\S_2^{\NG}(\rho) +\S_3^{\NG}(\rho) + \cdots\,.
\end{align}
Our aim in this paper is to compute the non-global functions $\S_m^{\NG}(\rho)$ for $m=2, 3, 4$ and $5$.

\subsection{Two-loops calculation and non-global logarithms}

The various squared-amplitudes for the emission of two energy-ordered (real or virtual) gluons with $x_2 \ll x_1 \ll 1$ are expressed as (see for instance \cite{Marchesini:1983bm}):
\begin{subequations}
\begin{align}
\W_{12}^{\R\R}=& \W_1^{\R}\W_2^{\R}+\oW^{\R\R}_{12}\,,
&\W_{12}^{\R\V}=-\W_{12}^{\R\R}\,, \label{eq:M12RR}\\
\W_{12}^{\V\R}=&-\W_1^{\R}\W_2^{\R}\,,
&\W_{12}^{\V\V}=-\W_{12}^{\V\R}\label{eq:M12VR}\,,
\end{align}
\end{subequations}
where the \emph{irreducible} term $\oW^{\R\R}_{12}$ is:
\begin{align}\label{eq:M12}
\oW^{\R\R}_{12} = & \, \frac{1}{2}\CF\CA \A_{q\bar{q}}^{12}\,,
\end{align}
and the measurement operator in this case is given by:
\begin{align}
\hat{\U}_2 = \hat{u}_1 \hat{u}_2 =& \left(1-\Theta^\rho_1\Theta^{\inn}_1\hat{\Theta}^\R_1\right) \left(1-\Theta^\rho_2\Theta^{\inn}_2\hat{\Theta}^\R_2\right)
\notag\\
 = &   1-\Theta^\rho_1\Theta^{\inn}_1\hat{\Theta}^\R_1- \Theta^\rho_1\Theta^\rho_2 \Theta^{\inn}_2\hat{\Theta}^\R_2\left(\hat{\Theta}^\V_1+\Theta^{\out}_1 \hat{\Theta}_1^\R \right),
\label{eq:u(k1k2)}
\end{align}
where $\Theta^\rho_2=\Theta^\rho_1\Theta^\rho_2$ since $x_1>x_2$. Acting on the squared amplitudes yields:
\begin{subequations}
\begin{align}
\hat{u}_1\hat{u}_2 \W_{12}^{\R\R}+\hat{u}_1\hat{u}_2 \W_{12}^{\R\V} =& -  \Theta^\rho_1\Theta^\rho_2 \Theta^{\inn}_2 \Theta^{\out}_1 \W_{12}^{\R\R}\,,
\\
\hat{u}_1\hat{u}_2 \W_{12}^{\V\R}+\hat{u}_1\hat{u}_2 \W_{12}^{\V\V} =& -  \Theta^\rho_1 \Theta^\rho_2\Theta^{\inn}_2 \W_{12}^{\V\R}\,.
\end{align}
\end{subequations}
The sum of these terms then gives:
\begin{align}
\sum_{\X} \hat{\U}_2\W_{12}^{\X} =&  - \Theta^\rho_1 \Theta^\rho_2 \Theta^{\inn}_2\left(\W_{12}^{\V\R}+\Theta^{\out}_1 \W_{12}^{\R\R}\right)
\notag\\
 = & - \Theta^\rho_1 \Theta^\rho_2 \Theta^{\inn}_2 \left(-\Theta^\inn_1\, \W_{1}^{\R}\W_2^\R+\Theta^{\out}_1\, \oW_{12}^{\R\R}\right),
\end{align}
where we used the squared amplitudes from eqs. \eqref{eq:M12RR} and \eqref{eq:M12VR}.
Substituting into the expression of $\S_2(\rho)$ we obtain:
\begin{align}\label{eq:Sigma2unintegrated}
\S_2(\rho) = & \int_{x_1>x_2}  \d\Phi_1 \Theta^\rho_1\Theta^{\inn}_1\, \W_1^{\R} \times \d\Phi_2 \Theta^\rho_2  \Theta^{\inn}_2\, \W_2^{\R}
 - \int_{x_1>x_2} \d\Pi_{12} \, \Theta^{\out}_1 \Theta^{\inn}_2\,  \oW_{12}^{\R\R}\,,
\end{align}
where we introduced the shorthand notation $\d\Pi_{12\cdots m}=\prod_{i=1}^m\d\Phi_i \Theta^\rho_i$. In the first integral in the right-hand-side of eq. \eqref{eq:Sigma2unintegrated}, the integrand is symmetric under the exchange $k_1\leftrightarrow k_2$,  which means that we can relax the condition $x_1>x_2$  and divide the integral by a factor 2!. Hence this integral factors out into the product of two separate identical contributions from gluons $k_1$ and $k_2$, which both have exactly the same form as the one-loop result (eq. \eqref{eq:S1}).
Thus we find for this term:
\begin{align}
\S_2^{\mathrm{P}}(\rho) =& \int_{x_1>x_2}  \d\Phi_1 \Theta^\rho_1\Theta^{\inn}_1\, \W_1^{\R} \times \d\Phi_2 \Theta^\rho_2  \Theta^{\inn}_2 \,\W_2^{\R}\notag\\
=&  \frac{1}{2!} \left(-\int  \d\Phi_1 \Theta^\rho_1\Theta^{\inn}_1\, \W_1^{\R}\right)^2 \notag\\
=&  \frac{1}{2!} \left(-\CF\bar{\alpha}_sL^2 \right)^2 = \frac{1}{2!} \left(\S_1^\mathrm{P}\right)^2,\label{eq:SP2}
\end{align}
which is just the expansion of the Sudakov $\S^\mathrm{P}(\rho)$ at second order. Hence:
\begin{subequations}
\begin{align}
\S_2(\rho) &= \S_2^\mathrm{P}(\rho) + \S_2^{\NG}(\rho)\,,\\
\S_2^{\NG}(\rho)& =
 - \int_{x_1>x_2} \d\Pi_{12} \Theta^{\out}_1\Theta^{\inn}_2\, \oW_{12}^{\R\R}\,.\label{eq:S2NG}
\end{align}
\end{subequations}
The latter expression is the pure non-global contribution at this order. It is given by:
\begin{equation}
\S_2^{\NG}(\rho) =  - \frac{1}{2} \CF\CA \bar{\alpha}_s^2 \int_{x_1>x_2} \frac{\d x_1}{x_1}  \frac{\d x_2}{x_2}\d \eta_1 \d \eta_2 \frac{\d\phi_1}{2\pi} \frac{\d\phi_2}{2\pi}  \Theta(x_2-\rho) \Theta(-\eta_1) \Theta(\eta_2)  \A_{q\bar{q}}^{12}\,,
\end{equation}
where we have $\Theta^\rho_2 = \Theta(x_2e^{-\eta_2}-\rho) \approx \Theta(x_2-\rho)$, since no collinear (double logarithms) are present for the pure non-global contribution. Hence the $x$ integration easily factors from the rapidity integration and we just set the lower limit on $x_2$ to $\rho$ (to single logarithmic accuracy).
Performing the trivial integration over $x_i$ we obtain the result $L^2/2!$. We note that at $n^{\mathrm{th}}$ order we have:
\begin{equation}
\int_\rho^1 \frac{\d x_1}{x_1} \int_\rho^{x_1} \frac{\d x_2}{x_2} \int_\rho^{x_2} \frac{\d x_3}{x_3} \cdots \int_\rho^{x_{n-1}} \frac{\d x_n}{x_n} = \frac{L^n}{n!}\,.
\end{equation}
Using the result of integration over $\phi_2$ and $\eta_2$ from eq. \eqref{eq:A4outq234} of appendix \ref{sec:azimuth} (with $\{j,m\} \to \{1,2\}$),  we obtain:
\begin{align}
\S_2^{\NG}(\rho) =&  - \frac{1}{2} \CF\CA \frac{\Lb^2}{2!} \int_{-\infty}^0 \d \eta_1 4\ln \frac{1}{1-e^{2\eta_1}}\notag\\
  =&  -\frac{1}{2}\CF \CA \frac{\Lb^2}{2!} \frac{\pi^2}{3}  = -\frac{\Lb^2}{2!}\CF \CA \zeta_2\,,\label{eq:Aqqbarint}
\end{align}
where $\Lb = \bar{\alpha}_s L$ and $\zeta$ is the Riemann-Zeta function. This is exactly the result obtained by Dasgupta and Salam \cite{Dasgupta:2001sh} for NGLs at two-loops. To the best of our knowledge the analytical calculation of NGLs at \emph{finite} $N_c$ beyond this order has not been performed before, and it is this very task that we do in the next section for the first time in the literature.

\section{Non-global logarithms beyond leading order}

\subsection{Three-loops calculation}

Having reproduced the well-known result for NGLs at leading order (two-loops), we proceed to compute NGLs at finite $N_c$ at next-to-leading order, namely triple gluons emission. As usual we begin by  the simplification of the measurement operator which will help us identify both angular and real-virtual configurations giving rise to large logarithms:
\begin{align}
\hat{\U}_3  =\hat{u}_1\hat{u}_2\hat{u}_3&= \left(1- \Theta^\rho_1 \Theta^{\inn}_1\hat{\Theta}^\R_1\right) \left(1- \Theta^\rho_2 \Theta^{\inn}_2\hat{\Theta}^\R_2\right) \left(1- \Theta^\rho_3 \Theta^{\inn}_3\hat{\Theta}^\R_3\right)\notag\\
&= \hat{\widetilde{\U}}_3- \Theta^\rho_1\Theta^\rho_2\Theta^\rho_3\Theta^{\inn}_3\hat{\Theta}^\R_3 \left(\hat{\Theta}^\V_2+\Theta^{\out}_2\hat{\Theta}^\R_2\right) \left(\hat{\Theta}^\V_1+\Theta^{\out}_1\hat{\Theta}^\R_1\right),
\end{align}
where $\hat{\widetilde{\U}}_3$ is the collection of terms which when acting on the squared amplitudes $\W_{123}^{\X}$, with $X$ summed over, yields a zero.
The action of the measurement operator on the various squared-amplitudes summed over $X$ gives:
\begin{equation}
\sum_{\X} \hat{\U}_3 \W_{123}^{\X} =
- \Theta^\rho_1\Theta^\rho_2\Theta^\rho_3\Theta^{\inn}_3  \left(\W_{123}^{\V\V\R}+\Theta^{\out}_2\, \W_{123}^{\V\R\R}+\Theta^{\out}_1\,\W_{123}^{\R\V\R}+\Theta^{\out}_1\Theta^{\out}_2\,\W_{123}^{\R\R\R}\right). \label{eq:U3MX3}
\end{equation}
As stated in the introduction, the explicit expressions for the various squared amplitudes above (together with those at higher loops) will be presented in our forthcoming work \cite{Delenda:eikamp}. Here we restrict ourselves to showing the simplification of the above squared-amplitudes in terms of the antenna functions defined previously in eq. \eqref{eq:antennas}. We have:
\begin{align}\label{eq:SUMX3}
\sum_{\X} \hat{\U}_3 \W_{123}^{\X} =- \Theta^\rho_1\Theta^\rho_2\Theta^\rho_3 \Theta^{\inn}_3
\times &\left(\Theta^{\inn}_1 \Theta^{\inn}_2\, \W_1^\R \W_2^\R \W_3^\R - \Theta^{\inn}_1 \Theta^{\out}_2\, \W_1^\R\oW_{23}^{\R\R} - \right.
\notag\\
&\quad-\Theta^{\out}_1 \Theta^{\inn}_2\, \W_2^\R \oW_{13}^{\R\R} - \Theta^{\out}_1 \Theta^{\inn}_2\, \W_3^\R \oW_{12}^{\R\R} + \notag\\
&\quad\left.+\Theta^{\out}_1\,\oW_{123}^{\R\V\R} + \Theta^{\out}_1 \Theta^{\out}_2\, \oW_{123}^{\R\R\R}\right).
\end{align}
In eq. \eqref{eq:SUMX3} $\W_i^\R$ and $\oW_{ij}^{\R\R}$ are defined at previous orders (see eqs. \eqref{eq:M1} and \eqref{eq:M12}), and $\oW_{123}^{\R\V\R}$ and $\oW_{123}^{\R\R\R}$ are the new irreducible terms of the squared amplitudes at this loop order proportional to the colour factor $\CF\CAsq$.

Thus the hemisphere mass distribution at $\mathcal{O}(\alpha_s^3)$, $\S_3(\rho)$, may be written as a sum of three contributions: $\S_3(\rho)=\S_3^A(\rho)+ \S_3^B(\rho) + \S_3^C(\rho)$. The first contribution is:
\begin{align}\label{eq:int3}
\S_3^A(\rho) =& - \int_{x_1>x_2>x_3} \d\Phi_1\Theta^\rho_1 \Theta^{\inn}_1\,\W_1^\R \times  \d\Phi_2\Theta^\rho_2\Theta^\inn_2\,\W_2^\R\times  \d\Phi_3\Theta^\rho_3\Theta^\inn_3 \,\W_3^\R \notag\\
=& \frac{1}{3!} \left(-\int  \d\Phi_1\Theta^\rho_1 \Theta^{\inn}_1  \,\W_1^\R\right)^3= \frac{1}{3!} \left(\S_1^\mathrm{P}\right)^3 = \S_3^\mathrm{P}(\rho)\,,
\end{align}
which is simply the expansion of the Sudakov $\S^\mathrm{P}(\rho)$ at this order. The factor $1/3!$ accounts for the fact that the integrand in eq. \eqref{eq:int3} is completely symmetric under the exchange of gluons, which means that the condition $x_1>x_2>x_3$ can be relaxed and the result multiplied by $1/3 !$. The integral is then factored out into the product of three identical integrals, each of them resembling the one-loop result eq. \eqref{eq:S1}.

The second contribution to $\S_3(\rho)$ is:
\begin{align}
\S_3^B(\rho) =& \int_{x_1>x_2>x_3}  \d\Pi_{123}\Theta^{\inn}_1\,\W_1^\R \Theta^{\out}_2 \Theta^{\inn}_3\,\oW_{23}^{\R\R}+ \int_{x_1>x_2>x_3}  \d\Pi_{123}\Theta^{\inn}_2 \,\W_2^\R\Theta^{\out}_1 \Theta^{\inn}_3\, \oW_{13}^{\R\R}+\notag\\
& + \int_{x_1>x_2>x_3}  \d\Pi_{123}\Theta^{\inn}_3\,\W_3^\R\Theta^{\out}_1 \Theta^{\inn}_2\,\oW_{12}^{\R\R}\,.\label{eq:S3B}
\end{align}
By swapping $k_1 \leftrightarrow k_2$ in the second integral of eq. \eqref{eq:S3B}, and performing the successive permutations: $k_1 \leftrightarrow k_2$ then $k_1 \leftrightarrow k_3$ in the third integral of the same equation, $\S_3^B (\rho)$ becomes:
\begin{align}
\S_3^B (\rho)=&\int_{x_1>x_2>x_3}  \d\Pi_{123} \Theta^{\inn}_1\,\W_1^\R \Theta^{\out}_2 \Theta^{\inn}_3 \,\oW_{23}^{\R\R} + \int_{x_2>x_1>x_3}  \d\Pi_{123} \Theta^{\inn}_1 \,\W_1^\R \Theta^{\out}_2 \Theta^{\inn}_3 \,\oW_{23}^{\R\R}+\notag\\
& + \int_{x_2>x_3>x_1}  \d\Pi_{123} \Theta^{\inn}_1 \,\W_1^\R \Theta^{\out}_2 \Theta^{\inn}_3 \,\oW_{23}^{\R\R}\,. \label{eq:S3BB}
\end{align}
The three integrands in eq. \eqref{eq:S3BB} are identical except for the region of integration over transverse momenta fractions. Thus we unify them into a single integral with the region of integration expressed by:
\begin{align}
\Theta(x_1-x_2) \Theta(x_2-x_3) + \Theta(x_2-x_1) \Theta(x_1-x_3) +  \Theta(x_2-x_3) \Theta(x_3-x_1) = \Theta(x_2-x_3)\,.
\end{align}
Hence we write:
\begin{align}
\S_3^B(\rho) =& \left(- \int \d\Phi_1 \Theta^\rho_1\Theta^{\inn}_1\,\W_1^\R\right) \times \left(-\int_{x_2>x_3}
\d\Pi_{23} \Theta^{\out}_2 \Theta^{\inn}_3 \oW_{23}^{\R\R}\right)\notag\\
=& \S_1^{\mr{P}}(\rho) \times \S_2^{\NG}(\rho)\,.
\end{align}
Thus $\S_3^B(\rho)$ factors out into a product of the one-loop primary cross-section and the two-loop NGLs cross-section (eqs. \eqref{eq:S1} and \eqref{eq:S2NG} respectively). This result is expected from the expansion of the Sudakov form factor at one-loop times the leading NGLs. Said differently, $\S_3^B(\rho)$ is just an ``interference'' term related to previous orders.

The remaining term $ \S_3^C(\rho)$, which is the pure irreducible NGLs contribution at this order, $\S_3^{\NG}(\rho)$, is proportional to $\CF\CAsq$ and given by:
\begin{align}\label{eq:S3G}
\S_3^{C}(\rho) =\S_3^{\NG}(\rho) =& - \int_{x_1>x_2>x_3} \d\Pi_{123} \Theta^{\out}_1 \Theta^{\inn}_3 \left(\oW_{123}^{\R\V\R}+\Theta^{\out}_2\,\oW_{123}^{\R\R\R}\right).
\end{align}
From the above equation one sees that the new irreducible NGLs contribution at three-loop order is generated by two mechanisms:
\begin{enumerate}[(a)]
\item the energy-ordered \emph{real} gluons $k_1$ and $k_2$ outside $\mathcal{H}_R$ coherently emit the softest gluon $k_3$ into $\mathcal{H}_R$ --- the term $\oW_{123}^{\R\R\R}$\,,
\item the hardest real gluon $k_1$ outside $\mathcal{H}_R$ emits the softest gluon $k_3$ inside, while $k_2$ is \emph{virtual} (inside or outside $\mathcal{H}_R$) --- the term $\oW_{123}^{\R\V\R}$.
\end{enumerate}
In both cases NGLs result from the miscancellation with the corresponding squared amplitude for $k_3$ virtual, i.e., the miscancellation  between $\oW_{123}^{\R\R\V}$ and $\oW_{123}^{\R\R\R}$ on the one hand, and between $\oW_{123}^{\R\V\V}$ and $\oW_{123}^{\R\V\R}$ on the other hand. Both of these contributions are not related to previous orders. It seems, at first inspection, that the second mechanism mentioned above (particularly the case where gluon $k_2$ is inside $\mathcal{H}_R$, as clearly shown in eq. \eqref{eq:S3_NG} below) is in contradiction with the common picture about the origin of NGLs. The latter picture dictates that, to all-orders, NGLs are entirely generated from a soft emission into $\mathcal{H}_R$ that is coherently radiated by arbitrary ensembles of soft, but harder, large-angle energy-ordered gluons outside $\mathcal{H}_R$ \cite{Dasgupta:2001sh, Dasgupta:2002bw}. Nonetheless, and though NGLs contribution from the said mechanism comes from both gluons $k_2$ and $k_3$ inside $\mathcal{H}_R$, gluon $k_2$ is actually virtual. We shall see later that this mechanism persists at higher loops too. Hence whenever a contribution to NGLs comes from configurations whereby gluons other than the softest are inside $\mathcal{H}_R$, then these ``other'' gluons must be virtual.\footnote{This observation was also made in ref. \cite{Schwartz:2014wha}.}

In fact the contributions of the two terms $\oW_{123}^{\R\V\R}$ and $\oW_{123}^{\R\R\R}$ in eq. \eqref{eq:S3G} are separately divergent but their sum is finite. The integral \eqref{eq:S3G} can be expressed as a sum of two finite terms in the following way:
\begin{align}
\S_3^{\NG}(\rho) =& - \int_{x_1>x_2>x_3} \d\Pi_{123} \Theta^{\out}_1 \Theta^{\inn}_3 \left(\Theta^{\inn}_2\, \oW_{123}^{\R\V\R}+\Theta^{\out}_2\left[\oW_{123}^{\R\V\R}+\oW_{123}^{\R\R\R}\right]\right).
\label{eq:S3_NG}
\end{align}
Substituting the explicit expressions of the irreducible terms $\oW_{123}^{\R\V\R}$ and $\oW_{123}^{\R\R\R}$ in terms of the antenna functions yields:
\begin{align} \label{eq:g3}
\S_3^{\NG}(\rho)
 =& \frac{1}{4}\CF\CAsq \,\frac{\Lb^3}{3!} \int_{-\infty}^0 d\eta_1\; 8 \ln^2 (1-e^{2\eta_1})-\notag\\
&-\frac{1}{4}\CF\CAsq \,\frac{\Lb^3}{3!} \int_{-\infty}^0 d\eta_1\; 2 \left(\A_{q1}^{\overline{23}}(\eta_1)+\A_{1\bar{q}}^{\overline{23}}(\eta_1)-2\zeta_2\right) ,
 \end{align}
where we performed the trivial integration over transverse momenta fractions to obtain $L^3/3!$ and used the results of rapidity and azimuthal integrations shown in appendix \ref{sec:azimuth}. The terms $\A_{q1}^{\overline{23}}(\eta_1)$ and $\A_{1\bar{q}}^{\overline{23}}(\eta_1)$ are given in eqs. \eqref{eq:further2} and \eqref{eq:further3} with $\{i,j,m\} \to \{1,2,3\}$.
The integration over $\eta_1$ in the first line of eq. \eqref{eq:g3} yields the result $8\zeta_3$ and that in the second line gives $4\zeta_3$.
Thus the pure non-global contribution at this order reads:
\begin{align}
\S_3^{\NG}(\rho)=&  \frac{\Lb^3 }{3!} \CF\CAsq\zeta_3\,.
\end{align}
Hence, up to this order we have:
\begin{align}\label{eq:S_NG_upto3loops}
\S^{\NG}(\rho)=&1-\frac{\Lb^2}{2!}\CF\CA  \zeta_2+\frac{\Lb^3}{3!}\CF\CAsq  \zeta_3+\mathcal{O}(\alpha_s^4)\,.
\end{align}

It is intriguing to note that the coefficients of NGLs at finite $N_c$ for two and three-loops (i.e., $\zeta_2$ and $\zeta_3$) are identical to those found at large $N_c$
\cite{Banfi:2002hw, Schwartz:2014wha}.\footnote{The appearance of $\zeta_2$ and $\zeta_3$ at two- and three-loop orders for NGLs is intriguing too. In appendix \ref{app:NoteOnZeta-NGLs} we present some useful observations regarding possible relations between the two (NGLs and Zeta function) seemingly distinct quantities.}  This is due to the fact that the combination of real/virtual squared amplitudes \emph{strangely} produces identical integrands in the expressions of the NGLs contributions $\S^{\NG}_{2}$ and $\S^{\NG}_3$ (eqs. \eqref{eq:S2NG} and \eqref{eq:S3_NG} respectively).
It would have been tremendously easy to resum NGLs for the hemisphere mass distribution to all-orders if the pattern in \eqref{eq:S_NG_upto3loops} persisted at higher loops. Although, we shall encounter Zeta functions at higher loop orders, the pattern itself unfortunately breaks down starting at four-loops, as we shall see in the next subsection.

\subsection{Four-loops calculation}

For four gluons emission, the computation of NGLs for the hemisphere mass distribution proceeds in an analogous manner to that of two and three gluons emission.
At this (four-loops) order the measurement operator reads:
\begin{align}
\hat{\U}_4&= \hat{u}_1\hat{u}_2\hat{u}_3 \hat{u}_4 = \left(1- \Theta^\rho_1 \Theta^{\inn}_1\hat{\Theta}^\R_1\right) \left(1- \Theta^\rho_2 \Theta^{\inn}_2\hat{\Theta}^\R_2\right) \left(1- \Theta^\rho_3 \Theta^{\inn}_3\hat{\Theta}^\R_3\right) \left(1- \Theta^\rho_4 \Theta^{\inn}_4\hat{\Theta}^\R_4\right)\notag\\
&= \hat{\widetilde{\U}}_4- \Theta^\rho_1\Theta^\rho_2\Theta^\rho_3\Theta^\rho_4\Theta^{\inn}_4\hat{\Theta}^\R_4  \left(\hat{\Theta}^\V_3+\Theta^{\out}_3\hat{\Theta}^\R_3\right) \left(\hat{\Theta}^\V_2+\Theta^{\out}_2\hat{\Theta}^\R_2\right) \left(\hat{\Theta}^\V_1+\Theta^{\out}_1\hat{\Theta}^\R_1\right),
\end{align}
where $\hat{\widetilde{\U}}_4$ is the sum of all terms which when operate on the squared amplitudes $\W_{1234}^{\X}$ and $X$ is summed over give zero. Acting by the measurement operator on the various squared amplitudes and summing over configurations we obtain:
\begin{align}
\sum_{\X} \hat{\U}_4 \W_{1234}^{\X} =
- \Theta^\rho_1\Theta^\rho_2\Theta^\rho_3\Theta^\rho_4\Theta^{\inn}_4
&\left(\W_{1234}^{\V\V\V\R} +\Theta^{\out}_1\, \W_{1234}^{\R\V\V\R} +\Theta^{\out}_2 \,\W_{1234}^{\V\R\V\R}+\right.\notag\\
&\left.\,+\Theta^{\out}_3\, \W_{1234}^{\V\V\R\R}+\Theta^{\out}_1 \Theta^{\out}_2\, \W_{1234}^{\R\R\V\R} +\Theta^{\out}_2 \Theta^{\out}_3\, \W_{1234}^{\V\R\R\R}+\right.\notag\\
&\left.\,+\Theta^{\out}_1 \Theta^{\out}_3\, \W_{1234}^{\R\V\R\R}+\Theta^{\out}_1 \Theta^{\out}_2 \Theta^{\out}_3\, \W_{1234}^{\R\R\R\R}\right). \label{eq:U4MX4}
\end{align}
From eqs. \eqref{eq:U3MX3} and \eqref{eq:U4MX4}, it should be clear how the result of the action of the measurement operator on the squared amplitudes at $m^{\mr{th}}$ loop order would look like:
\begin{itemize}
\item the softest gluon is always inside $\mathcal{H}_R$\,,
\item each \emph{real} gluon $k_i$ is associated with a step function $\Theta^{\out}_i$\,,
\item virtual gluons are associated with neither $\Theta^{\inn}$ nor $\Theta^{\out}$  (i.e., they can either be in or out of $\mathcal{H}_R$).
\end{itemize}

The hemisphere mass distribution at fourth order may then be cast in the form \eqref{eq:Sigma_m}.
Substituting the various matrix-elements squared we can split the hemisphere mass distribution, $\S_{4}(\rho)$, into five parts:
$\S_4=\S_4^A+\S_4^B+\S_4^C+\S_4^D +\S_4^E$, with:
\begin{align}
\S_4^A =& \int_{x_1>x_2>x_3>x_4} \d\Pi_{1234} \, \Theta^{\inn}_1\Theta^{\inn}_2\Theta^{\inn}_3\Theta^{\inn}_4\, \W^\R_1\W^\R_2 \W^\R_3 \W^\R_4\,,
\notag \\
\S_4^B =&-\int_{x_1>x_2>x_3>x_4} \d\Pi_{1234} \Big( \Theta^{\out}_1 \Theta^{\inn}_2\Theta^{\inn}_3\Theta^{\inn}_4\, \oW^{\R\R}_{12}\W^{\R}_3\W^\R_4
+2 \leftrightarrow 3+2 \leftrightarrow 4+ \notag\\
&\qquad \qquad\qquad\qquad+[1 \leftrightarrow 3\,\textrm{and} \,2 \leftrightarrow 4]+[2 \leftrightarrow 1\,\textrm{then}\,1 \leftrightarrow 4] + [2 \leftrightarrow 1\,\textrm{then}\,1 \leftrightarrow 3]\Big)\,,
\notag\\
\S_4^C = & \int_{x_1>x_2>x_3>x_4} \d\Pi_{1234} \,\Big\{\Theta^{\out}_1 \Theta^{\inn}_3\Theta^{\inn}_4\, \W_4^\R\left(\oW^{\R\V\R}_{123}+\Theta^{\out}_2\,\oW^{\R\R\R}_{123}\right)+ 3\leftrightarrow 4+\notag\\
 &\qquad \qquad\qquad\qquad\qquad +[3\leftrightarrow 2 \,\textrm{then}\,2\leftrightarrow 4]+ [1\leftrightarrow 2 \,\textrm{then}\,1\leftrightarrow 3\,\textrm{then}\,1\leftrightarrow 4 ]\Big\}\,,
 \notag\\
\S_4^D =& \int_{x_1>x_2>x_3>x_4} \d\Pi_{1234} \, \left(\Theta^{\out}_1 \Theta^{\inn}_2\Theta^{\out}_3\Theta^{\inn}_4\, \oW^{\R\R}_{12}\oW^{\R\R}_{34} + 2\leftrightarrow 3 + 2\leftrightarrow 3\,\textrm{then} \, 3\leftrightarrow 4\right),
\notag\\
\S_4^E =& -\int_{x_1>x_2>x_3>x_4} \d\Pi_{1234} \, \Theta^{\out}_1  \Theta^{\inn}_4\times\notag\\
&\qquad \times\left(\oW^{\R\V\V\R}_{1234}+\Theta^{\out}_3\, \oW^{\R\V\R\R}_{1234}+\Theta^{\out}_2\,   \oW^{\R\R\V\R}_{1234} + \Theta^{\out}_2\Theta^{\out}_3\, \oW^{\R\R\R\R}_{1234}\right),
\end{align}
where terms in the last line are the four-loop irreducible components of the squared amplitudes for the corresponding gluon configurations. The parts $\S_4^A$,  $\S_4^B$, $\S_4^C$, and $\S_4^D$ completely reduce to integrals we calculated at previous orders, while the remaining $\S_4^E$ part is the new NGLs contribution. Let us evaluate each of these integrals separately starting with the reducible parts.

\subsubsection{Reducible parts}

For the first part $\S_4^A$ we can, as usual, relax the condition $x_1>x_2>x_3>x_4$ and multiply the result by a factor of $1/4!$. This part then factors out
into the product of four identical integrals of the form we met at $\mathcal{O}(\alpha_s)$ (eqs. \eqref{eq:S1} and \eqref{eq:S1_b}), thus we obtain for this term:
\begin{align}
\S_4^A(\rho) =& \frac{1}{4!} \left( -\int \d\Phi_1 \Theta^\rho_1 \Theta^{\inn}_1\, \W_1^\R  \right)^4= \frac{1}{4!} \left(\S_1^\mathrm{P}\right)^4 = \S_4^\mathrm{P}(\rho)\,,
\end{align}
which is just the expansion of the Sudakov at the fourth order.

The second part $\S_4^B$ is carried out in a fashion analogous to that of $\S_3^B$ in eq. \eqref{eq:S3B}. The five integrands are transformed into the first integrand, $\oW^{\R\R}_{12} \W^\R_3 \W^\R_4$, with the appropriate changes in the integration limits. Thus we have six integrals having identical integrands but different regions of integrations. Writing these integration regions as step functions and simplifying we obtain $\Theta(x_1-x_2)\Theta(x_3-x_4)$. This means that $\S_4^B$ factors out into the product of two integrals, one over $k_1$ and $k_2$ and the other over $k_3$ and $k_4$, as follows:
\begin{align}
\S_4^B(\rho) =& \left(\int_{x_3>x_4} \d\Phi_3  \Theta^\rho_3 \Theta^{\inn}_3\, \W^{\R}_3 \;\times \d\Phi_4 \Theta^\rho_4  \Theta^{\inn}_4\,  \W^\R_4\right)
 \times \left(- \int_{x_1>x_2} \d\Pi_{12} \Theta^{\out}_1 \Theta^{\inn}_2\, \oW^{\R\R}_{12}\right)\notag\\
= & \frac{1}{2!}(\S_1^\mathrm{P})^2 \times\S_2^{\NG}, \label{eq:S4B}
\end{align}
where we have used eqs. \eqref{eq:SP2} and \eqref{eq:S2NG} to arrive at the second line of the above equation. Eq. \eqref{eq:S4B} is in fact the interference of the expansion of the Sudakov form factor $\S^{\mr{P}}$ with NGLs at two-loops $\S_2^{\NG}$.

Performing the integrations in the third part $\S_4^C$ along the same lines outlined above for $\S^B_4$ (i.e., by making appropriate changes of variables) we obtain:
\begin{align}
\S_4^C(\rho) =& \left(-\int  \d\Phi_4 \Theta^\rho_4\Theta^{\inn}_4 \,\W^\R_4\right) \times  \left(-\int_{x_1>x_2>x_3} \d\Pi_{123} \Theta^{\out}_1 \Theta^{\inn}_3  \left[\oW^{\R\V\R}_{123}+\Theta^{\out}_2\oW^{\R\R\R}_{123}\right]\right)
\notag\\
= & \S_1^\mathrm{P}(\rho) \times\S_3^{\NG}(\rho)\,,
\end{align}
where we used eqs. \eqref{eq:S1} and \eqref{eq:S3G}. This result is the interference between the expansion of the Sudakov and $\S_3^{\NG}$.

The part $\S_4^D$ can be written as follows:
\begin{align}\label{eq:S4D}
\S_4^D(\rho) =& \frac{1}{2} \int_{x_1>x_2} \d\Pi_{12} \Theta^{\out}_1 \Theta^{\inn}_2 \,  \oW^{\R\R}_{12}\times
\int_{x_3>x_4}  \d\Pi_{34} \Theta^{\out}_3\Theta^{\inn}_4\, \oW^{\R\R}_{34}\notag\\
=& \frac{1}{2}  \left(\S_2^{\NG}\right)^2,
\end{align}
which indicates a possible pattern of exponentiation of NGLs since this term resembles the structure of the expansion of $\exp\{\S_2^{\NG}\}$ at this (fourth) order.

\subsubsection{Irreducible part}

The irreducible part at four-loops, $\S_4^{E}(\rho)$, is given by:
\begin{align}
\S_4^E (\rho) =& -\int_{x_1>x_2>x_3>x_4} \d\Pi_{1234} \, \Theta^{\out}_1  \Theta^{\inn}_4\times\notag\\
& \times\left(\oW^{\R\V\V\R}_{1234}+\Theta^{\out}_3 \, \oW^{\R\V\R\R}_{1234}+\Theta^{\out}_2\, \oW^{\R\R\V\R}_{1234} + \Theta^{\out}_2\Theta^{\out}_3 \, \oW^{\R\R\R\R}_{1234}\right). \label{eq:S4E}
\end{align}
The first irreducible squared-amplitudes $\oW_{1234}^{\R\V\V\R}$ and $\oW_{1234}^{\R\V\R\R}$ are proportional to $\CF\CAcub$, whereas the other two amplitudes, $\oW_{1234}^{\R\R\V\R}$ and $\oW_{1234}^{\R\R\R\R}$, contain both $\CF\CAcub$ and $\CFsq\CAsq$ terms. The phase space integration of all terms in eq.~\eqref{eq:S4E} breaks the simple pattern observed in eq. \eqref{eq:S_NG_upto3loops}, with the last two terms ($\oW_{1234}^{\R\R\V\R}$ and $\oW_{1234}^{\R\R\R\R}$) even breaking the colour pattern.
Therefore the loss of the pattern of NGLs is in fact a manifestation of the break in the structure of the eikonal amplitudes. This might be related to the failure of the ``probabilistic scheme'' discussed by Dokshitzer et al. in ref. \cite{Dokshitzer:1991wu}, where such (irreducible) contributions to the eikonal squared-amplitude were dubbed ``monster'' terms. They were traced back to be originating from the ``colour polarisability'' of  jets \cite{Dokshitzer:1991wu}.\footnote{More details are to be found in our forthcoming paper \cite{Delenda:eikamp}.} Moreover, we note here, as can be seen from the form of $\S_4^E$, that contributions to NGLs at this order are generated when the softest gluon $k_4$ is emitted inside the measured hemisphere $\mathcal{H}_R$, whilst the hardest gluon $k_1$ is always outside. The other two gluons, $k_2$ and $k_3$, may be emitted inside $\mathcal{H}_R$ provided they are virtual (in accordance with the observation made in the previous subsection).

The contribution $\S^E_{4}$ is not related to the expansion of the Sudakov nor to NGLs at previous orders, and represents the \emph{new} non-global contribution at four-loops. Together with $\S_4^D$ they form the total non-global contribution, $\S_4^{\NG}$, to the hemisphere mass distribution at this order. To evaluate the part $\S_4^E$ we first integrate over transverse momenta fractions, which as usual yields $L^4/4!$, and then perform the azimuthal and rapidity integrations according to the angular configurations indicated in eq. \eqref{eq:S4E}. The azimuthal integrations are not as straightforward as at three-loops and we find the method used in ref. \cite{Schwartz:2014wha} of contour integration very useful in reducing the number of integrals to be performed. The reduced integrals are then carried out analytically whenever possible, otherwise numerically. Since all analytical integrations yield results that are explicitly proportional to $\zeta_4$, the resultant values from numerical integrations were interpreted in terms of $\zeta_4$. In addition to this semi-numerical approach we also verify our results by numerically integrating each of the finite terms in the following form of $\S_4^E (\rho)$:
\begin{align}
 \S_4^E (\rho) =& -\int_{x_1>x_2>x_3>x_4} \d\Pi_{1234} \, \Theta^{\out}_1  \Theta^{\inn}_4\times\notag\\
&\times\Big\{\Theta^{\inn}_2\Theta^{\inn}_3 \,\oW^{\R\V\V\R}_{1234}
+\notag\\
&\qquad +\Theta^{\inn}_2 \Theta^{\out}_3 \left(\oW^{\R\V\V\R}_{1234} + \oW^{\R\V\R\R}_{1234} \right) + \Theta^{\out}_2 \Theta^{\inn}_3 \left(\oW^{\R\V\V\R}_{1234} + \oW^{\R\R\V\R}_{1234} \right)
+ \notag\\
&\qquad + \Theta^{\out}_2 \Theta^{\out}_3 \left(\oW^{\R\V\V\R}_{1234} + \oW^{\R\V\R\R}_{1234} + \oW^{\R\R\V\R}_{1234} + \oW^{\R\R\R\R}_{1234} \right)
\Big\}\,,
\label{eq:S4E_finite}
\end{align}
over the \emph{full} (7-dimensional) phase space using the multi-dimensional numerical-integration library {\tt Cuba} \cite{Hahn:2004fe}.
The final result reads:
\begin{align}
\S_4^E(\rho) = - \frac{\Lb^4}{4!} \left(\frac{25}{8}\CF\CAcub+\CFsq \CAsq\right)\zeta_4\,,
\end{align}
which may also be rewritten in the following two alternative forms:
\begin{subequations}
\begin{align}
 \S_4^E(\rho)
 &= - \frac{\Lb^4}{4!}\,\CF\CAcub\,\zeta_4\,\left[\frac{29}{8} + \left(\frac{\CF}{\CA} - \frac{1}{2} \right) \right] \label{eq:S4_finite_nc_corr}
 \\
 &= - \frac{\Lb^4}{4!} \left[\frac{25}{8}\,\CF\CAcub\,\zeta_4 + \frac{2}{5}\, \left(\CF\CA\,\zeta_2\right)^2 \right].\label{eq:S4_pattern}
\end{align}
\end{subequations}
The expression \eqref{eq:S4_finite_nc_corr} explicitly shows the finite-$N_c$ correction to the large-$N_c$ result, while \eqref{eq:S4_pattern} emphasises the pattern $\CF\CAn\,\zeta_{\mr{n}+1}$ seen at two and three-loops. It also reveals that even though $\S^E_4$ is a new irreducible contribution at four-loops, it still contains factors related to lower-order NGL contributions (the term $(\CF\CA\,\zeta_2)^2$). Observe that the size of the finite-$N_c$ correction in \eqref{eq:S4_finite_nc_corr} is about $\sim 1.5\%$ that of the large-$N_c$ result. This is in agreement with the conclusion arrived at in \cite{Hatta} for the impact of finite-$N_c$ corrections at all-orders for $e^+e^-$ processes. The total non-global contribution at this order is then given by:
\begin{align}\label{eq:S4NG_total}
\S_4^\NG(\rho) = \S_4^D+\S_4^E &= -
\frac{\Lb^4}{4!}  \left(\frac{25}{8}\, \CF\CAcub \zeta_4 - \frac{13}{5}\,\CFsq\CAsq\,\zeta_2^2 \right),
\end{align}
and thus the hemisphere mass distribution up to this order is expressed as:
\begin{align}\label{eq:S4Fin}
\S(\rho)=&\S^{\mathrm{P}}(\rho)\times \S^{\NG}(\rho)\,,
\\
\S^{\NG}(\rho) =& 1-\frac{\Lb^2}{2!} \CF\CA \zeta_2  + \frac{\Lb^3}{3!} \CF\CAsq \zeta_3  - \frac{\Lb^4}{4!} \left[  \frac{25}{8} \CF\CAcub \zeta_4 - \frac{13}{5}\,\CFsq\CAsq\,\zeta_2^2 \right] +\mathcal{O}(\alpha_s^5)\,.\notag
\end{align}
In the next subsection we discuss the five-loops case and the possibility of resummation of NGLs.

\subsection{Five-loops and beyond}

Following the same steps as before we write the measurement operator at five-loops as follows:
\begin{align}
\hat{\U}_5=& \left(1- \Theta^\rho_1 \Theta^{\inn}_1\hat{\Theta}^\R_1\right) \left(1- \Theta^\rho_2 \Theta^{\inn}_2\hat{\Theta}^\R_2\right) \left(1- \Theta^\rho_3 \Theta^{\inn}_3\hat{\Theta}^\R_3\right) \left(1- \Theta^\rho_4 \Theta^{\inn}_4\hat{\Theta}^\R_4\right)\left(1- \Theta^\rho_5 \Theta^{\inn}_5\hat{\Theta}^\R_5\right)\notag\\
=& \hat{\widetilde{\U}}_5- \Theta^\rho_1\Theta^\rho_2\Theta^\rho_3\Theta^\rho_4\Theta^\rho_5\Theta^{\inn}_5\hat{\Theta}^\R_5
  \left(\hat{\Theta}^\V_4+\Theta^{\out}_4\hat{\Theta}^\R_4\right) \left(\hat{\Theta}^\V_3+\Theta^{\out}_3\hat{\Theta}^\R_3\right)\left(\hat{\Theta}^\V_2+\Theta^{\out}_2\hat{\Theta}^\R_2\right)\times \notag\\
 &\times \left(\hat{\Theta}^\V_1+\Theta^{\out}_1\hat{\Theta}^\R_1\right),
\end{align}
where, as usual, $\hat{\widetilde{\U}}_5$ is the sum of all terms that yield vanishing contributions to the hemisphere mass distribution. Acting by the measurement operator on the various squared amplitudes and summing over configurations we obtain:
\begin{align}\label{eq:EikAmp5}
\sum_{\X}& \hat{\U}_5 \W_{12345}^{\X} = - \prod_{i=1}^5 \Theta^\rho_i \,\Theta^{\inn}_5\times \notag\\
&\left(\W_{12345}^{\V\V\V\V\R} +\Theta^{\out}_1 \W_{12345}^{\R\V\V\V\R} +\Theta^{\out}_2 \W_{12345}^{\V\R\V\V\R}+\Theta^{\out}_3 \W_{12345}^{\V\V\R\V\R}
+\Theta^{\out}_4 \W_{12345}^{\V\V\V\R\R}+\right.\notag\\
&+\Theta^{\out}_1 \Theta^{\out}_2 \W_{12345}^{\R\R\V\V\R}+\Theta^{\out}_1 \Theta^{\out}_3 \W_{12345}^{\R\V\R\V\R}+\Theta^{\out}_1 \Theta^{\out}_4 \W_{12345}^{\R\V\V\R\R}+\Theta^{\out}_2 \Theta^{\out}_3 \W_{12345}^{\V\R\R\V\R}+\notag\\
& +\Theta^{\out}_2 \Theta^{\out}_4 \W_{12345}^{\V\R\V\R\R}+\Theta^{\out}_3 \Theta^{\out}_4 \W_{12345}^{\V\V\R\R\R}+\Theta^{\out}_1 \Theta^{\out}_2\Theta^{\out}_3 \W_{12345}^{\R\R\R\V\R}+
\notag\\
&+\Theta^{\out}_1 \Theta^{\out}_2\Theta^{\out}_4 \W_{12345}^{\R\R\V\R\R}+\Theta^{\out}_1 \Theta^{\out}_3\Theta^{\out}_4 \W_{12345}^{\R\V\R\R\R}+\Theta^{\out}_2 \Theta^{\out}_3\Theta^{\out}_4 \W_{12345}^{\V\R\R\R\R}+\notag\\
&\left.+\Theta^{\out}_1 \Theta^{\out}_2\Theta^{\out}_3\Theta^{\out}_4 \W_{12345}^{\R\R\R\R\R}\right).
\end{align}
The hemisphere mass distribution at five-loops is then given in eq. \eqref{eq:Sigma_m} with $m=5$. Substituting the various matrix-elements squared and following the same procedure outlined at four-loops we again obtain two types of contributions; reducible, $\S^{\mr{r}}_5$, and irreducible, $\S^{\mr{irr}}_5$. The former contains all the interference terms between the Sudakov factor $\S^{\mr{P}}$ and NGLs at previous orders as well as interference terms between two and three-loops NGLs. Explicitly written it reads:
\begin{align}\label{eq:S5_red}
\S^{\mr{r}}_5(\rho) =& \frac{1}{5!} \left( \S^{\mr{P}}_1 \right)^5 + \frac{1}{3!} \left( \S^{\mr{P}}_1 \right)^3 \times \S^{\NG}_2 + \frac{1}{2!} \left( \S^{\mr{P}}_1 \right)^2 \times \S^{\NG}_3 +
\S^{\mr{P}}_1 \times \S^{\NG}_4 + \S^{\NG}_2 \times \S^{\NG}_3\,.
\end{align}
Note that the penultimate term in the above equation contains the contribution $\S^{\mr{P}}_1 \times (\S^{\NG}_2)^2/2!$ (eqs. \eqref{eq:S4NG_total} and \eqref{eq:S4D}).

The irreducible contribution $\S^{\mr{irr}}_5$ is expressed as:
\begin{align}\label{eq:S5irr}
\S_5^\irr =&  - \int_{x_1>x_2>x_3>x_4>x_5} \d\Pi_{12345} \,\Theta^{\out}_1\Theta^{\inn}_5\times \notag\\
&\times \left( \oW_{12345}^{\R\V\V\V\R}+\Theta^{\out}_2 \oW_{12345}^{\R\R\V\V\R}+ \Theta^{\out}_3 \oW_{12345}^{\R\V\R\V\R}+ \Theta^{\out}_4 \oW_{12345}^{\R\V\V\R\R} + \Theta^{\out}_2\Theta^{\out}_3 \oW_{12345}^{\R\R\R\V\R}+\right.\notag\\
&\qquad\left.+ \Theta^{\out}_2\Theta^{\out}_4 \oW_{12345}^{\R\R\V\R\R}+\Theta^{\out}_3\Theta^{\out}_4 \oW_{12345}^{\R\V\R\R\R}+ \Theta^{\out}_2\Theta^{\out}_3\Theta^{\out}_4 \oW_{12345}^{\R\R\R\R\R}\right),
\end{align}
which is neither related to the Sudakov factor nor to NGLs at previous orders, and which contains both $\CFsq \CAcub$ and $\CF\CAfour$ terms. The irreducible squared-amplitudes that contribute to $\S_5^\irr$ can be classified into two types:
\begin{itemize}
\item proportional only to $\CF \CAfour$: $\oW_{12345}^{\R\V\R\R\R}$, $\oW_{12345}^{\R\V\R\V\R}$, $\oW_{12345}^{\R\V\V\R\R}$, and $\oW_{12345}^{\R\V\V\V\R}$,
\item containing both $\CFsq\CAcub$ and $\CF \CAfour$: $\oW_{12345}^{\R\R\R\R\R}$, $\oW_{12345}^{\R\R\R\V\R}$, $\oW_{12345}^{\R\R\V\R\R}$, and $\oW_{12345}^{\R\R\V\V\R}$, which all contain ``monster'' terms.
\end{itemize}
The calculation of $\S_5^\irr$ turns out to be trickier and more involved than anticipated. In particular we have not yet been able to simplify (like we did at four-loops) the monster parts of the irreducible amplitudes of the second type above, to forms that can readily be integrated. Other than the monster parts, all remaining terms (either of the first or second type above) are in fact integrable.

Till the full expression of the irreducible contribution is simplified and integrated, we express the result of  $\S_5^{\irr}$ in the following form (based on the pattern seen at four-loops \eqref{eq:S4_pattern} and the pieces found in the integrable amplitudes of eq. \eqref{eq:S5irr}):
\begin{subequations}
\begin{align}
\S_5^{\irr} &= \frac{\Lb^5}{5!}\,\CF\CAfour\,\zeta_5 \left[\alpha + \beta \left(\frac{\CF}{\CA} - \frac{1}{2} \right) \right] \label{eq:S5irr_finiteNc}
\\
 &= \frac{\Lb^5}{5!} \left[\left(\alpha - \frac{\beta}{2}\right) \CF\CAfour\,\zeta_5 + a \beta\, \CFsq\CAcub\, \zeta_2\zeta_3 \right], \label{eq:S5irr_pattern}
\end{align}
\end{subequations}
where $a = \zeta_5/\zeta_2 \zeta_3 \simeq 0.5244 $ and the constant coefficients $\alpha$ and $\beta$ are yet to be determined. We discuss the possible values of these constants when we compare our results with those at large $N_c$ in the next section. The form \eqref{eq:S5irr_finiteNc} explicitly shows the finite-$N_c$ correction. The total NGLs contribution at five-loops then reads:
\begin{align}\label{eq:sigma5NG}
\S_5^{\NG}=& \S_2^{\rm{NG}} \times \S_3^{\NG}+\S_5^{\irr}\notag\\
=& - \frac{\Lb^5}{2! 3!}\,\CFsq\CAcub\,\zeta_2 \zeta_3 + \S_5^{\irr}\notag\\
=& \frac{\Lb^5}{5!} \left[\left(\alpha-\frac{\beta}{2}\right)\CF\CAfour\,\zeta_5 - (10 - a \beta)\,\CFsq\CAcub\,\zeta_2\zeta_3 \right].
\end{align}

The results we obtained up to five-loops, particularly eq. \eqref{eq:S5_red}, in fact suggest a \emph{possible} resummation of NGLs into an exponential function of the form:
\begin{align}\label{eq:SNG_resummed5loops}
\S^{\NG}(\rho) = \exp \bigg\{ & -\frac{\Lb^2}{2!} \CF\CA \zeta_2  + \frac{\Lb^3}{3!} \CF\CAsq \zeta_3  - \frac{ \Lb^4}{4!}\, \CF\CAcub\,\zeta_4 \left[\frac{29}{8} +  \left(\frac{\CF}{\CA}-\frac{1}{2}\right) \right] +\notag\\
& +\frac{\Lb^5}{5!}\,\CF\CAfour\,\zeta_5 \left[\alpha + \beta \left(\frac{\CF}{\CA}-\frac{1}{2}\right) \right] + \mathcal{O} (\alpha_s^6)\bigg\}\,.
\end{align}
Eq. \eqref{eq:SNG_resummed5loops} may actually be rewritten in a form analogous to that found in ref. \cite{Delenda:2012mm} (eqs. (5.10) and (5.11)) for clustering logarithms. To this end we write:
\begin{align}\label{eq:SNG_CLs_analog}
 \S^{\NG}(\rho) = \exp\left[ -\frac{\CF}{\CA} \sum_{n\geq2} \frac{1}{n!}\,\cS_n\, \left(-\CA\,\Lb\right)^n \right],
\end{align}
where
\begin{equation}
\cS_2 = \zeta_2\,,\quad \cS_3 = \zeta_3\,,\quad \cS_4 = \zeta_4 \left[\frac{29}{8} + \left(\frac{\CF}{\CA} - \frac{1}{2} \right)\right], \quad \cS_5 = \zeta_5 \left[\alpha + \beta \left(\frac{\CF}{\CA} - \frac{1}{2} \right)\right].
\end{equation}
The above-mentioned similarity between clustering logarithms and NGLs emphasises the common (non-global) origin of the two types of logarithms.
Moreover, following the pattern in \eqref{eq:S5irr_pattern}, eq. \eqref{eq:SNG_resummed5loops} may also be recast into the form:
\begin{align}\label{eq:SNG_resummed5loops_pattern}
\S^{\NG}(\rho)  = \exp \bigg\{ &-\frac{\Lb^2}{2!} \CF\CA \zeta_2  + \frac{\Lb^3}{3!} \CF\CAsq \zeta_3 - \frac{ \Lb^4}{4!}  \left[\frac{25}{8} \CF\CAcub\,\zeta_4 + \frac{2}{5}\, \CFsq\CAsq\,\zeta_2^2 \right] +\notag\\
& + \frac{\Lb^5}{5!} \left[\left(\alpha - \frac{\beta}{2}\right) \CF\CAfour\,\zeta_5 + a \beta\, \CFsq\CAcub\,\zeta_2\zeta_3 \right] + \mathcal{O} (\alpha_s^6)\bigg\}\,.
\end{align}
The expansion of the above exponential exactly reproduces the terms we have calculated up to five-loops including all interference terms in the distribution. At each higher order one simply adds a new irreducible NGLs term in the exponent.

In fact, if the pattern deduced in eqs. \eqref{eq:S4_finite_nc_corr} and \eqref{eq:S4_pattern} persists at higher-loop orders, then one can put forth the following \emph{ansatz} for the general form of the $n^{\mr{th}}$ order contribution to the exponent of the resummed NGLs factor:
\begin{subequations}\label{eq:Sn_ansatz}
\begin{align}
& (-1)^{n-1}\,\frac{\Lb^n}{n!}\,\CF\CAnm\,\zeta_{n} \left[ \gamma_n + \sum_{k=2}^{\lfloor n/2 \rfloor} \sigma_k \left( \left[\frac{\CF}{\CA}\right]^{k-1} - \frac{1}{2^{k-1}} \right) \right],
\\
& (-1)^{n-1}\,\frac{\Lb^n}{n!} \left[\bar{\gamma}_n\,\CF\CAnm\,\zeta_n + \sum_{k=2}^{\lfloor n/2\rfloor} \bar{\sigma}_k\,\CF^{\!\!\!\!k}\,\CA^{\!\!\!\!n-k}\,\zeta_k \zeta_{n-k} \right],
\end{align}
\end{subequations}
where $\lfloor n\rfloor$ represents the floor function of $n$, and $\gamma_n$, $\sigma_n$, $\bar{\gamma}_n$, and $\bar{\sigma}_k$ are constant coefficients to be determined from integrations. The two formulae presented above for the ansatz are equivalent up to the constant coefficients. The first form stresses the finite-$N_c$ correction while the second preserves the pattern seen at two-, three- and four-loops (eq. \eqref{eq:S4_pattern}). The above formulae may only be verified once higher-loop orders are carried out explicitly. We hope to perform such calculations in the near future.

\begin{figure}
\centering
\input{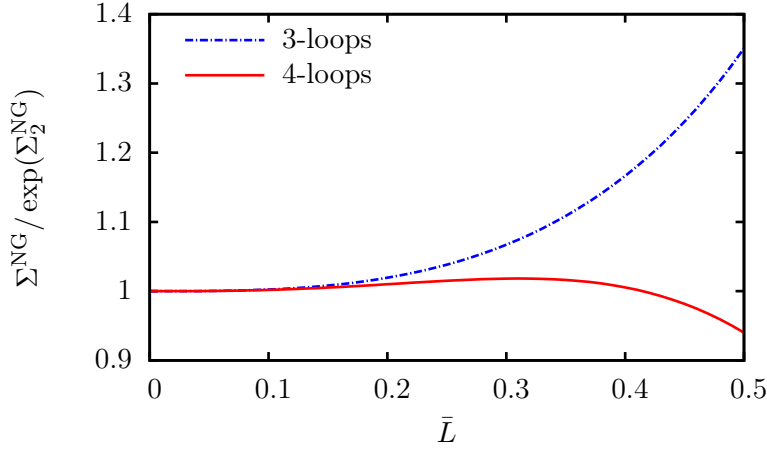}
\caption{\label{fig:loops} Plot of the ratio $\S^{\NG}(\rho)/\exp(\S_2^{\NG}(\rho))$ in terms of the logarithm $ \Lb= \as/\pi \ln(1/\rho)$.}
\end{figure}

We note that in the exponent of \eqref{eq:SNG_resummed5loops} (and \eqref{eq:SNG_resummed5loops_pattern}) the series in $\Lb$ has alternating signs at each escalating order.
To assess the relative size of the three and four-loops corrections to the leading two-loops result, we plot in figure \ref{fig:loops} the ratio $\S^{\NG}/\exp(\S^{\NG}_2)$ for various truncations of the series in the exponent in eq. \eqref{eq:SNG_resummed5loops}. The leading NGLs coefficient seems to dominate for only relatively small values of $\Lb$ ($\Lb \lesssim 0.15$). For larger values the series seems to depart from the leading term in an alternating way (towards larger (smaller) values for odd (even) loop orders).
These significant variations mean that the terms computed thus far are insufficient to capture the full behaviour of the all-orders resummed distribution. We expect, however, that adding few more terms in the exponent may lead to a convergent and more stable behaviour, since one could argue that higher-order terms are suppressed by $\Lb^n/n!$, while $\zeta_n$ saturates at $1$ as $n$ becomes larger.

To cross-check our results, eqs. \eqref{eq:SNG_resummed5loops} and \eqref{eq:SNG_resummed5loops_pattern}, we compare them, in the next section, to previous calculations at large $N_c$ both at fixed order and to all-orders.

\section[Comparison with large-\texorpdfstring{$\mbox{\boldmath{$N_c$}}$}{Nc} results]{\boldmath Comparison with large-$N_c$ results}

\subsection[Comparison with analytic results at large \texorpdfstring{$N_c$}{Nc}]{\boldmath Comparison with analytical results at large $N_c$}

Having calculated the coefficients of NGLs at finite $N_c$ fully up to four-loops and partially at five-loops, we can now compare our findings to those of Schwartz and Zhu \cite{Schwartz:2014wha} obtained through the analytical solution to the BMS equation \cite{Banfi:2002hw} in the large-$N_c$ limit. To go from the finite-$N_c$ case to the large-$N_c$ approximation we simply invoke the replacement $\CF\to \CA/2 = N_c/2$ (where $\CA = N_c$). This is equivalent to expanding $\CF$ to first order in colour:
\begin{align}
  \CF &= \frac{N_c^2-1}{2N_c} = \frac{N_c}{2}+\mathcal{O}\left(\frac{1}{N_c}\right).
\end{align}
Then the expansion of our result \eqref{eq:SNG_resummed5loops_pattern} to leading order in colour, i.e., at large $N_c$, up to five-loops is:
\begin{align}\label{eq:S4largeNc}
\S^{\NG}= & 1-\frac{\pi^2}{24} (N_c\bar{ L})^2+\frac{\zeta_3}{12} (N_c\Lb)^3 + \frac{\pi^4}{34\,560} (N_c\Lb)^4+\notag\\
&+ \left(-\frac{\pi^2 \zeta_3}{288} + \frac{(2\alpha-\beta) \zeta_5 + a \beta \frac{\pi^2}{6} \zeta_3}{480}\right) (N_c\Lb)^5
+ \mathcal{O}\left((N_c \Lb)^6\right)\,,
\end{align}
where we have written the explicit values of $\zeta_2 = \pi^2/6$ and $\zeta_4 = \pi^4/90$.
The full result reported by Schwartz and Zhu (SZ) at large $N_c$, $\S^{\NG}_{\mr{SZ}}$,  is \cite{Schwartz:2014wha}:
\begin{align}\label{eq:S4largeNcSchwartz}
\S_{\mr{SZ}}^{\NG}=&
1-\frac{\pi^2}{24} \Lh^2+\frac{\zeta_3}{12} \Lh^3 + \frac{\pi^4}{34\,560} \Lh^4+\left(-\frac{\pi^2 \zeta_3}{360}+\frac{17}{480}\zeta_5\right) \Lh^5+\mathcal{O}(\alpha_s^6)\,,
\end{align}
where $\Lh$ is simply $N_c\,\Lb$. The two results are thus identical up to four-loops. Recalling that they were arrived at using different approaches, their equality provides a solid cross-check of the correctness of the computed NGLs coefficients (at least up to four-loops). Even though we are unable to fully compare our result to that of Schwartz and Zhu at five-loops, due to the missing values of $\alpha$ and $\beta$, it is ironic to note that the pattern spotted at four-loops, eq. \eqref{eq:S4_pattern}, seems to hold true at five-loops. Whilst the term $\zeta_5$ is apparent in \eqref{eq:S4largeNcSchwartz}, the product $\zeta_2 \zeta_3$ is disguising in the factor $\pi^2 \zeta_3/360$. A quick comparison reveals the values:
\begin{align}
\alpha = \frac{17}{2} + \frac{1}{a}\,, \qquad \beta = \frac{2}{a}\,.
\end{align}

Given the above values we expect the finite-$N_c$ result of NGLs up to five-loops to be expressed as:
\begin{align}\label{eq:SNG_resummed+full5loops}
\S^{\NG}(\rho) = \exp
\bigg\{ &-\frac{\Lb^2}{2!} \CF\CA \zeta_2  + \frac{\Lb^3}{3!} \CF\CAsq \zeta_3  - \frac{ \Lb^4}{4!}  \left[\frac{25}{8}\,\CF\CAcub\,\zeta_4 + \frac{2}{5}\,\CFsq\CAsq\,\zeta_2^2 \right] +\notag\\
&+ \frac{\Lb^5}{5!} \left[\frac{17}{2}\,\CF\CAfour\,\zeta_5 + 2\,\CFsq\CAcub\,\zeta_2\zeta_3 \right] + \mathcal{O} (\alpha_s^6)\bigg\}\,.
\end{align}
We are not, however, claiming to have fully accounted for NGLs at this order since we have not explicitly calculated the coefficients of NGLs with colour factors $\CFsq\CAcub$ and $\CF\CAfour$. We hope that further research on this can help verify the above equation (and eq. \eqref{eq:Sn_ansatz} in general) in which case it may actually be possible to find a key to the analytical resummation of NGLs both at large and finite $N_c$.

Our approach additionally has the benefit that it sheds light on the possibility of assessing the validity of the large-$N_c$ approximation, by means of judging the impact of neglected finite-$N_c$ corrections. An important note in this regard is that at two- and three-loops, as can be seen by comparing eqs. \eqref{eq:S4Fin} and \eqref{eq:S4largeNc} at $\mathcal{O}(\Lb^2)$ and $\mathcal{O}(\Lb^3)$, there are no \emph{hidden} terms buried by the large-$N_c$ approximation, and the finite-$N_c$ result can simply be obtained from the solution of the BMS equation by just restoring the full colour factors through: $N_c^2 \to 2\CF\CA$ (at two-loops) and $N_c^3 \to 2\,\CF\CAsq$ (at three-loops). At four-loops this is not true and in fact there is a hidden correction that is given plainly in \eqref{eq:S4_finite_nc_corr}. We regard this as the first-order \emph{proper} finite-$N_c$ correction which introduces \emph{new} terms that are entirely absent at large $N_c$. The second-order proper finite-$N_c$ correction occurs at five-loops and is shown in \eqref{eq:S5irr_finiteNc}.

\subsection{Comparison with all-orders numerical results}

In order to verify our resummed formula \eqref{eq:SNG_resummed5loops}, and even \eqref{eq:SNG_resummed+full5loops} which includes the five-loops term, it is instructive to compare it to the all-orders numerical solution of the finite-$N_c$ Weigert equation \cite{Weigert:2003mm}. We have not, unfortunately, been able to obtain the output of the MC program, written by Hatta and Ueda \cite{Hatta:2013iba}, for the hemisphere mass distribution.\footnote{The hemisphere mass distribution has not yet been coded into the MC program \cite{Hatta}.} We thus postpone this discussion till the said numerical distribution becomes available. Furthermore, to assess the importance of the missing higher-loop terms in \eqref{eq:SNG_resummed5loops} we compare it to either the results obtained by the numerical solution of the BMS equation \cite{Banfi:2002hw} or to the output of the numerical MC program of Dasgupta and Salam (DS) \cite{Dasgupta:2001sh}. The latter two numerical solutions are in fact identical within a percent accuracy \cite{Banfi:2002hw, Schwartz:2014wha}, and we thus restrict ourselves to the DS MC program.

Let us introduce the standard evolution parameter $t$ \cite{Dasgupta:2001sh}, which accounts for the running of the coupling:
\begin{align}\label{eq:evol}
  t =& \frac{1}{2\pi} \int_{e^{-L}}^1 \alpha_s(Q x) \frac{dx}{x} = \frac{1}{4\pi \beta_0}\ln\frac{1}{1-2\beta_0 \alpha_s L}\,,
\end{align}
where $\beta_0$ is the one-loop coefficient of the QCD $\beta$ function. At fixed order one has $t = \as\,L/2\pi = \Lb/2$. Hence substituting $\Lb$ by $2\,t$ into eq. \eqref{eq:SNG_resummed5loops} up to four-loops we find:
\begin{align}\label{eq:prop}
\S^{\NG}(t) =&
 \exp\left( -\CF\CA \frac{\pi^2}{3}  \, t^2  + \frac{4}{3}\CF\CAsq \zeta_3 \,  t^3 -\frac{\pi^4}{135}  \left[ \frac{25}{8} \CF\CAcub + \CFsq\CAsq\right]  t^4 +\mathcal{O}(t^5) \right).
\end{align}
We compare the result \eqref{eq:prop} with the parametrisation for NGLs to all-orders obtained in ref. \cite{Dasgupta:2001sh} by fitting to the output of the aforementioned DS MC program \cite{Dasgupta:2001sh}:
\begin{align}\label{eq:DS_resumFactor}
\S_{\mr{DS}}^{\NG}(t) =&
 \exp\left( -\CF \CA \frac{\pi^2}{3} \frac{1+(0.85\CA t)^2}{1+(0.86 \CA t)^{1.33}}\,t^2 \right).
\end{align}

\begin{figure}[t]
\centering
\includegraphics[width=0.49\textwidth]{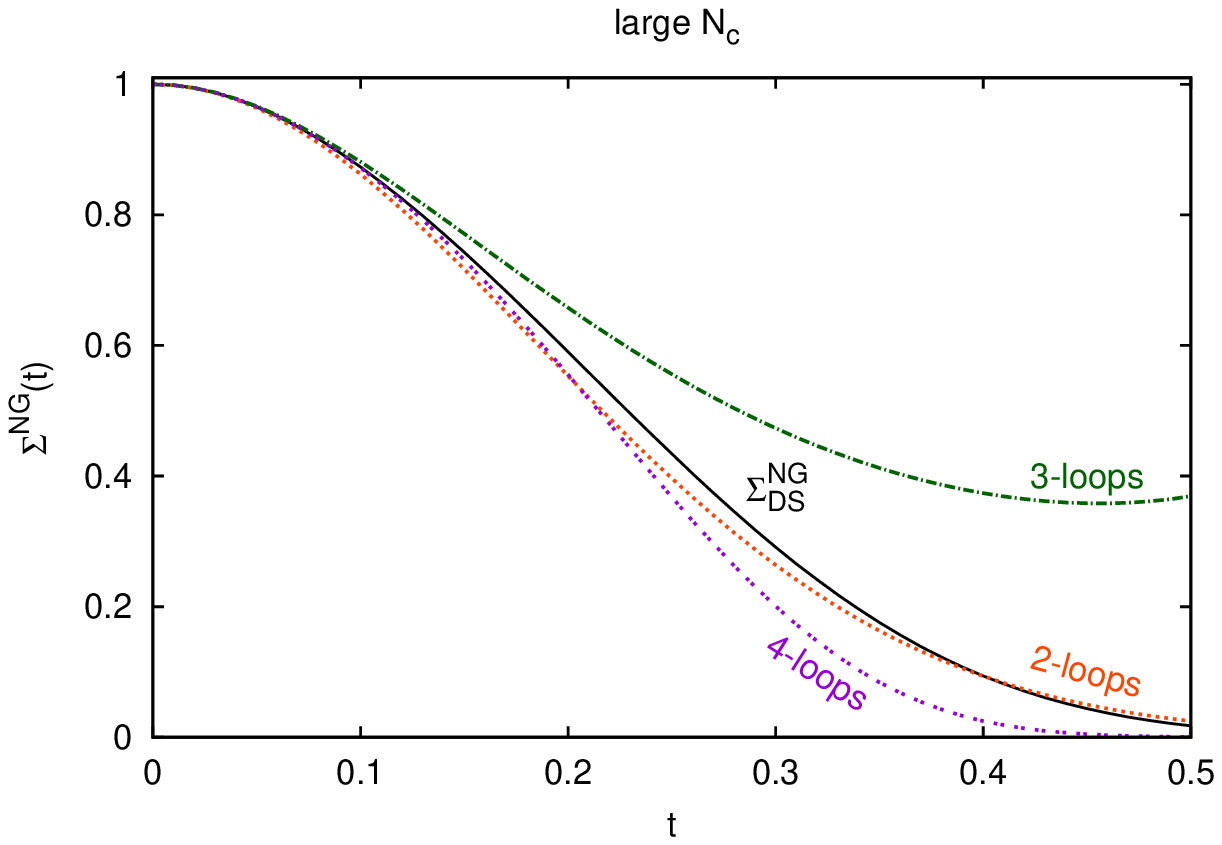}\hfill
\includegraphics[width=0.49\textwidth]{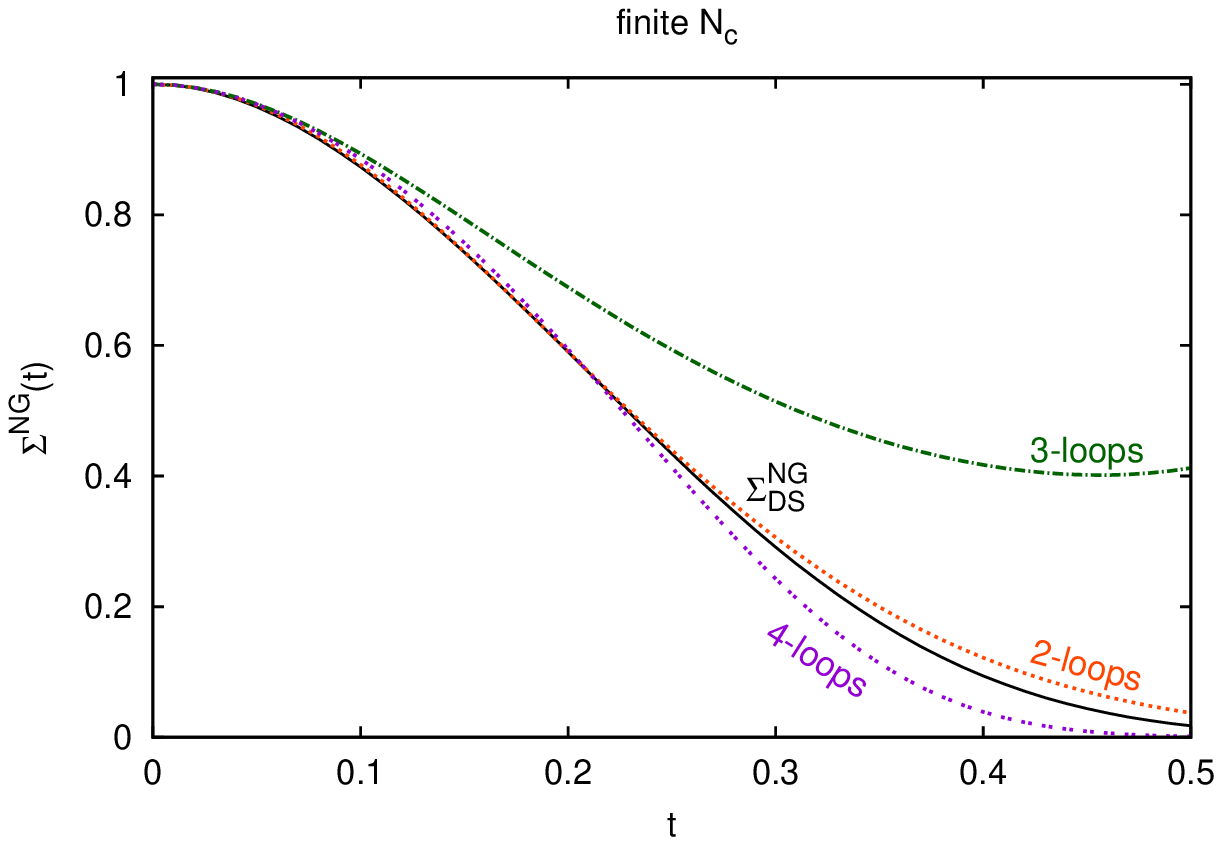}
\caption{\label{fig:SNG_4loops} Plot of the NGLs function $\S^{\NG}(\rho)$ at large (left) and finite (right) $N_c$.}
\end{figure}
In figure \ref{fig:SNG_4loops} we plot our approximate resummed result \eqref{eq:prop} for various truncations along with the DS resummed factor \eqref{eq:DS_resumFactor} for a range of $t \in [0,0.5]$ both at large (left) and finite (right) $N_c$. Recall that a value of $t  = 0.3$ corresponds to a value of $L=19$ and $\rho \sim 10^{-8}$ for $\alpha_s \sim 0.1$, which is sufficient for phenomenological purposes. Few points to note from the plots. Firstly, as expected for finite $N_c$, all curves are shifted up due to the fact that one is using $\CF = 4/3 \simeq 1.33$ instead of $\CF=\CA/2 = 3/2 = 1.5$ (recall that $\CF$ is in the exponent). Secondly, it is striking to observe that the best approximation to the all-orders result for quite a large range of $t$ is the leading two-loops result,\footnote{This observation was made in refs. \cite{Rubin:2010fc, Schwartz:2014wha} too.} $\exp\left(\S^\NG_2 \right)$, for both large and finite-$N_c$ cases. This suggests that the alternating, positive and negative, higher-loop contributions to $\S^\NG$ somehow balance out.

\begin{figure}
\centering
\includegraphics[width=0.49\textwidth]{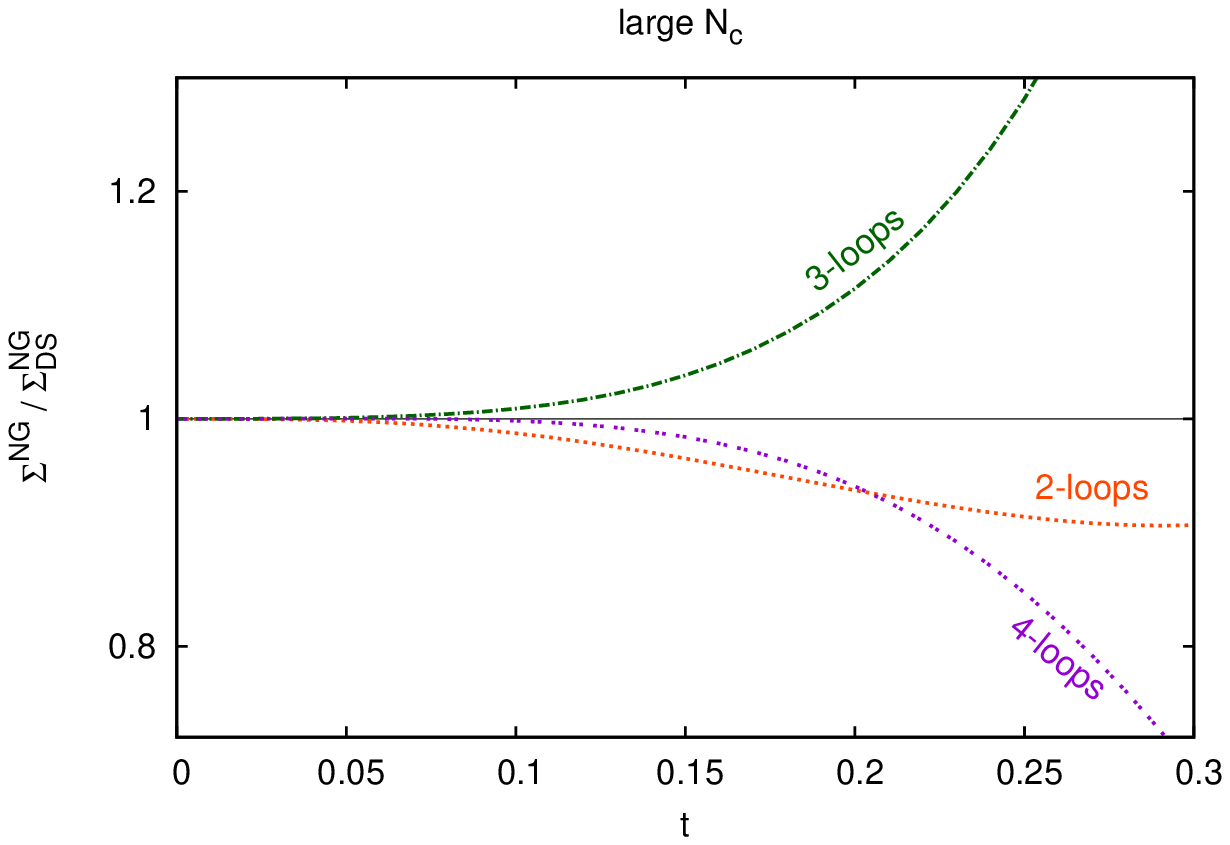}\hfill
\includegraphics[width=0.49\textwidth]{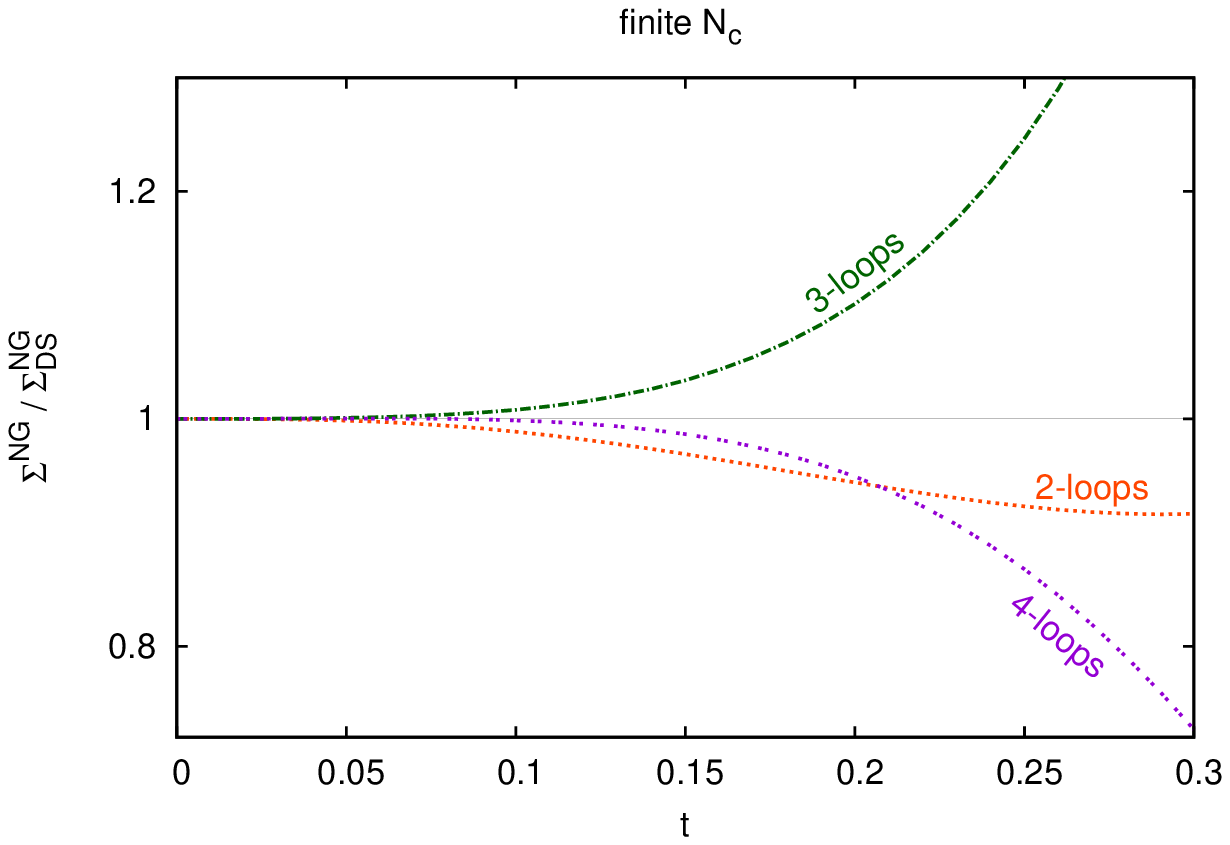}
\caption{\label{fig:SNG_ratio_4loops} Plot of the ratio $\S^{\NG}(\rho)/\S_{\mr{DS}}^{\NG}(\rho)$ for both large (left) and finite (right) $N_c$.}
\end{figure}

\begin{figure}
\centering
\includegraphics[width=0.7\textwidth]{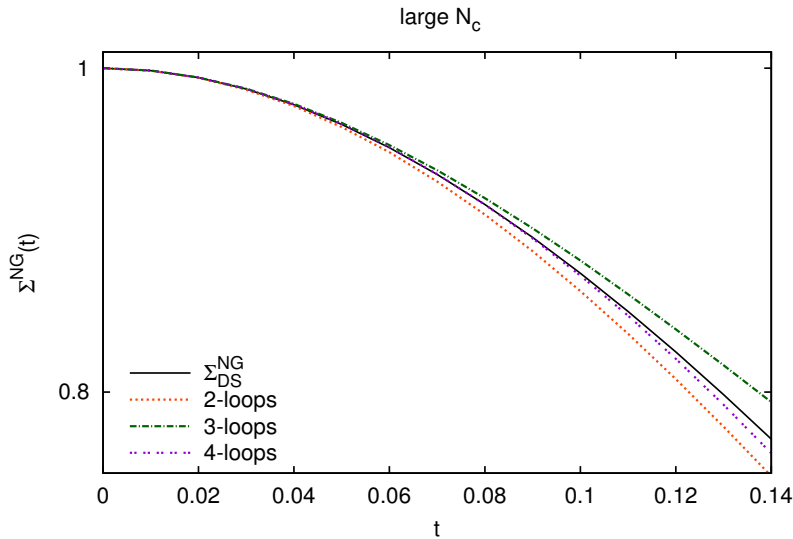}
\caption{\label{fig:SNG_4loops_t02} Plot of the NGLs function $\S^{\NG}(t)$ at large $N_c$ for the range $0 \leq t \leq 0.14$.}
\end{figure}

Moreover, the main feature of the plots and which has direct link to the purpose of this paper is actually seen at small values of $t$. We see that the interval of $t$ over which the four-loops result and the all-orders resummed factor \emph{overlap} is $0 \leq t \lesssim 0.12$. This interval of overlapping is smaller for three-loops, $ 0\leq t\lesssim 0.08$, and even smallest for two-loops $0 \leq t \lesssim 0.05$. The latter feature may be seen more clearly in figures \ref{fig:SNG_ratio_4loops} and \ref{fig:SNG_4loops_t02}. One would therefore expect that adding more terms in the exponent of \eqref{eq:SNG_resummed5loops} leads to increasingly larger intervals of overlapping.

A similar observation was also made in ref. \cite{Rubin:2010fc} for ``filtering analysis'' in the case of the filtering parameters $n_{\mathrm{filt}} = 2$ and $\eta_{\mathrm{filt}}=0.1, 0.3$ (figure 18 of ref. \cite{Rubin:2010fc}), as well as $n_{\mathrm{filt}} =3$ and $\eta_{\mathrm{filt}}=0.3$ (figure 21). However the author of ref. \cite{Rubin:2010fc}, having plotted the expansion of the full filtered Higgs-jet mass distribution including both primary and non-global logarithms in the Cambridge-Aachen jet algorithm \cite{Dokshitzer:1997in, Wobisch:1998wt}, ascribed the convergence of the series, as one adds higher-loop terms, to the dominance of the primary series. The author then verified this explanation by plotting the same distribution for higher values of $\eta_{\mathrm{filt}}$ (figure 19) where collinear logarithms are expected to be absent and NGLs become of the same order as primary logarithms. Two issues to point out regarding our work compared to that of ref. \cite{Rubin:2010fc}: firstly, we are plotting purely the NGLs resummed exponential factor and hence the convergence seen in figures \ref{fig:SNG_ratio_4loops} and \ref{fig:SNG_4loops_t02} has nothing to do with primary logarithms. Secondly, it is well known \cite{Appleby:2002ke, Banfi:2005gj, Delenda:2006nf, Banfi:2010pa, KhelifaKerfa:2011zu} that employing the Cambridge-Aachen jet algorithm not only reduces the size of NGLs but also introduces clustering logarithms that are as important as NGLs. Thus plotting the full distribution, which includes primary, non-global and clustering logarithms, would not tell much about the convergence of the NGLs series.

Moreover, the author of ref. \cite{Rubin:2010fc} also plotted (figure 24) the pure NGLs resummed factor for the interjet energy flow distribution and concluded that, up to six-loops, the NGLs series seems to be divergent. Recalling that the coefficient of the two-loops NGLs depends on the rapidity gap $\Delta\eta$ \cite{Dasgupta:2002bw, KhelifaKerfa:2011zu}, it is likely that higher-loop NGLs coefficients depend on $\Delta\eta$ too. The divergence may thus be due to the presence of the $\Delta\eta$ terms. For the hemisphere mass distribution that we have treated in this paper there is no such rapidity gap dependence. A proper answer, however, may only be given once the former --- interjet energy flow --- distribution is carefully considered, a task which we hope to perform in coming publications.

Notice finally that the so far discussed NGLs behaviour is in contrast to that seen for clustering logarithms \cite{Delenda:2012mm} where the whole structure of the all-orders result is mostly captured by the first few terms in the exponent.

% \begin{figure}
% \centering
% \includegraphics[width=0.7\textwidth]{Figs/Sng_DS_5l_t03_nc}
% \caption{\label{fig:SNG_5loops_t03} Plot of the NGLs function $\S^{\NG}(t)$ up to five-loops at large $N_c$ for the range $0 \leq t \leq 0.14$.}
% \end{figure}

\section{Conclusions}

In this paper we have considered the calculation of the leading NGLs at finite $N_c$ up to five-loops for the hemisphere mass distribution in $e^+ e^- \to$ di-jet events. We performed the calculation by means of integrating the squared amplitudes for the emission of energy-ordered soft gluons in the eikonal approximation, valid at single logarithmic accuracy, over a suitable phase space achieved through a measurement operator. The two and three-loops results were shown to be relatively straightforward to obtain, and were found to be directly related to the Riemann-Zeta function. We noticed that, up to this loop order, finite-$N_c$ corrections are absent. This was a direct consequence of the relatively simple structure of the eikonal amplitudes (up to this order) as well as the combination of real/virtual amplitudes induced by the phase space measurement operator.

Within the same eikonal framework we computed the four-loops contribution to NGLs distribution. The latter turns out to be much harder than the previous two orders and the simple result in terms of a product of a single colour factor and a Zeta function breaks down. This failure originates from a break in the simple structure of the corresponding real-virtual eikonal amplitudes at four-loops order, a phenomenon that was noticed more than two decays ago \cite{Dokshitzer:1991wu}. Nevertheless, we were able to overcome this complexity, compute NGLs and even spot a new pattern for NGLs at and beyond this order. This pattern helped us to successfully write down the five-loops contribution to NGLs up to constant coefficients which we extracted from comparisons to previous large-$N_c$ results \cite{Schwartz:2014wha}. We hope to be able to fully compute these constants in the near future. The five-loops calculation reveals that the NGLs distribution seems to exhibit a pattern of exponentiation. To this end the said distribution was cast in an exponential form with full NGLs coefficients and colour factors up to four-loops in the exponent.

Comparisons to large-$N_c$ results obtained by other authors \cite{Schwartz:2014wha} confirmed our findings, at least in the latter limit. We then took the step forth and compared our exponential function to the all-orders numerical resummed result reported by Dasgupta and Salam \cite{Dasgupta:2001sh}. To our surprise, the shape of the all-orders result was best represented by the two-loops approximation for a wide range of the evolution parameter $t$. In the region of small $t$, however, adding more terms in the exponent of our resummed result yielded better agreement, then the two-loops result, with the all-orders numerical result. This suggests that more higher-loop contributions are needed for our result to be of any phenomenological significance (i.e., till the agreement extends to values of $t$ up to $\sim$ 0.2 -- 0.3). The task of computing these higher-loop terms might not be impossible after all given that we have developed, over the course of preparing this paper, the machinery for: computing eikonal amplitudes at finite $N_c$ to theoretically any loop order, reducing the dimension of the phase space over which to integrate, and spotting a pattern for NGLs at each order.

\acknowledgments

We would like to thank Prof. Abdelhamid Bouldjedri, head of PRIMALAB research laboratory at Batna University, for making at our service the computer resources of the laboratory. We would also like to thank Mrinal Dasgupta for his comments on the manuscript. This work is supported in part by CNEPRU Research Project D01320130009.

\appendix

\section{Angular integrations}
\label{sec:azimuth}

In this section we present some definitions and azimuthal/rapidity integrations which have proven useful in our calculations.

Following ref. \cite{Schwartz:2014wha} we define the following angular functions:
\begin{subequations}
\begin{align}
(ij)=& \cosh(\eta_i-\eta_j)-\cos(\phi_i-\phi_j)\,,\\
\langle ij \rangle = &\frac{(ij)}{2\sinh\eta_i\sinh\eta_j}\,.
\end{align}
\end{subequations}
The basic antennas are expressed as:
\begin{align}
w_{q\bar{q}}^1 = 2 \,, \qquad
w_{q1}^2 = \frac{e^{-\eta_1+\eta_2}}{(12)}\,, \qquad
w_{1\bar{q}}^2 = \frac{e^{\eta_1-\eta_2}}{(12)}\,,\qquad
w_{12}^3 = \frac{(12)}{(13)(23)}\,.
\end{align}
The $\phi_i$-azimuthal averaging over the inverse of the angular function $(ij)$ is given by:
\begin{align}
\int_0^{2\pi} \frac{d\phi_i}{2\pi}\frac{1}{(ij)} = \csch|\eta_i-\eta_j|\,.
\end{align}
In the case where a gluon $k_m$ is constrained within the measured hemisphere region and gluons $k_i$ and $k_j$ are constrained outside, the $k_m$ angular integration over $w_{ij}^m$ yields  \cite{Schwartz:2014wha}:
\begin{align}
\int_0^\infty d\eta_m \int_0^{2\pi} \frac{d\phi_m}{2\pi} w_{ij}^m = \ln (1+\langle ij\rangle ) = \ln \frac{\cosh(\eta_i+\eta_j)-\cos(\phi_{i}-\phi_j)}{2\sinh\eta_i\sinh\eta_j}\,,
\end{align}
with $\eta_i<0$ and $\eta_j<0$. Furthermore, the azimuthal average over the angle $\phi_i$ of the emitter $k_i$ yields \cite{Schwartz:2014wha}:
\begin{align}
\int_0^{2\pi} \frac{d\phi_i}{2\pi}\frac{1}{(ij)} \ln (1+\langle ij\rangle)=\csch(\eta_i-\eta_j)\ln \frac{1-\coth\eta_i}{1-\coth\eta_j}\,.
\end{align}

With the same conditions ($\eta_i<0$, $\eta_j<0$ and $\eta_m>0$) we can perform the azimuthal and rapidity integrations for the following antenna functions:
\begin{subequations}
\begin{align}
\mathcal{A}_{q\bar{q}}^{j\overline{m}}\equiv
\int_0^\infty d\eta_m \int_0^{2\pi} \frac{d\phi_m}{2\pi} \mathcal{A}_{q\bar{q}}^{jm} =& -4 \ln(1-e^{2\eta_j})\,,\label{eq:A4outq234}\\
\mathcal{A}_{qi}^{j\overline{m}}\equiv
\int_0^\infty d\eta_m \int_0^{2\pi} \frac{d\phi_m}{2\pi} \mathcal{A}_{qi}^{jm} =&w_{qi}^j \ln \frac{1-\coth\eta_j}{1-\coth\eta_i}\frac{\cosh(\eta_j+\eta_i)-\cos(\phi_j-\phi_i)}{2\sinh\eta_j\sinh\eta_i}\, ,\\
\mathcal{A}_{i\bar{q}}^{j\overline{m}}\equiv
\int_0^\infty d\eta_m \int_0^{2\pi} \frac{d\phi_m}{2\pi} \mathcal{A}_{i\bar{q}}^{jm} =&w_{i\bar{q}}^j \ln \frac{1-\coth\eta_j}{1-\coth\eta_i}\frac{\cosh(\eta_j+\eta_i)-\cos(\phi_j-\phi_i)}{2\sinh\eta_j\sinh\eta_i}\,.
\end{align}
\end{subequations}
We can also perform further integrations over $\phi_j$ and $\eta_j$:
\begin{subequations}
\begin{align}
\mathcal{A}_{q\bar{q}}^{\overline{jm}}=& \int_{-\infty}^0 d\eta_j \int_0^{2\pi} \frac{d\phi_j}{2\pi} \mathcal{A}_{q\bar{q}}^{j\overline{m}}
 = \frac{\pi^2}{3} = 2\zeta_2 \,,\label{eq:further1}\\
\mathcal{A}_{qi}^{\overline{jm}}=& \int_{-\infty}^0 d\eta_j \int_0^{2\pi} \frac{d\phi_j}{2\pi} \mathcal{A}_{qi}^{j\overline{m}}
\notag\\
=&\ln (1-\tanh\eta_i) \ln ((\coth\eta_i-1)\coth\eta_i)+2\Li_2\frac{1}{1-\tanh\eta_i}-2\Li_2 \tanh\eta_i\,,\label{eq:further2}\\
\mathcal{A}_{i\bar{q}}^{\overline{jm}} =& \int_{-\infty}^0 d\eta_j \int_0^{2\pi} \frac{d\phi_j}{2\pi} \mathcal{A}_{i\bar{q}}^{j\overline{m}}\notag\\
=&\ln^2 2-\frac{\pi^2}{2}+ 2\ln (1-\coth\eta_i)\ln \frac{1-\coth\eta_i}{2}+2\ln (-\tanh\eta_i)\ln (-\csch(2\eta_i))+\notag\\
&+2\Li_2\frac{1}{1-\tanh\eta_i}+2\Li_2\frac{1-\tanh\eta_i}{2}+2\Li_2 (1+\tanh\eta_i)\,,\label{eq:further3}
\end{align}
\end{subequations}
with $\Li_2$ the polylogarithm function of order 2.
Similarly integrating over the angles of the softest particle $k_n$ in the antenna $\mathcal{A}_{ij}^{mn}$ yields:
\begin{align}
\mathcal{A}_{ij}^{m\overline{n}}=& w_{ij}^m \left( \ln(1+\langle im\rangle) + \ln(1+\langle jm\rangle) - \ln(1+\langle ij\rangle) \right).
\end{align}

At four-loops, the following azimuthal integrations are relevant:
\begin{align}
\int_0^{2\pi} \frac{1}{(13)(23)}\frac{d\phi_3}{2\pi}=&
\frac{\coth|\eta_1-\eta_3|}{2\sinh(\eta_1-\eta_3)\sinh(\eta_2-\eta_3) } \times \\
& \times \left(\frac{\sinh|\eta_1-\eta_3| \csch|\eta_2-\eta_3|-\cosh(\eta_1-\eta_2)}
{\cosh(\eta_1-\eta_2)-\cos(\phi_1-\phi_2)}+\right.\notag\\
&\qquad\left.+\frac{\cosh(\eta_1+\eta_2-2\eta_3)-\sinh|\eta_1-\eta_3| \csch|\eta_2-\eta_3|}{\cosh(\eta_1+\eta_2-2\eta_3)-\cos(\phi_1-\phi_2)}
\right)+\eta_1\leftrightarrow \eta_2\,.\notag
\end{align}
Then we have:
\begin{align}
& \int_0^{2\pi} \frac{d\phi_2}{2\pi}\int_0^{2\pi} \frac{d\phi_3}{2\pi} \frac{1}{(13)(23)} \ln(1+\langle 12\rangle)
=\frac{\coth|\eta_1-\eta_3|}
{2\sinh(\eta_1-\eta_3)\sinh(\eta_2-\eta_3) } \times \\
& \times \left(\frac{\sinh|\eta_1-\eta_3| \csch|\eta_2-\eta_3|-\cosh(\eta_1-\eta_2)}
{\sinh(\eta_1-\eta_2)} \ln\frac{\coth\eta_1-1}{\coth\eta_2-1} +\right.\notag\\
&\qquad+\frac{\cosh(\eta_1+\eta_2-2\eta_3)-\sinh|\eta_1-\eta_3| \csch|\eta_2-\eta_3|}{\sinh|\eta_1+\eta_2-2\eta_3|}\times\notag\\
&\qquad\times\left.\left\{\ln \left[\csch\eta_1 \csch\eta_2 \sinh ^2\frac{\eta_1+\eta_2-|\eta_1+\eta_2-2\eta_3|}{2}\right]-|\eta_1+\eta_2-2\eta_3|\right\}\right)+\eta_1\leftrightarrow \eta_2\,. \notag
\end{align}

\section[A note on NGLs-\texorpdfstring{$\mbox{\boldmath{$\zeta_n$}}$}{zeta\_n} relation]{\boldmath A note on NGLs-$\zeta_n$ relation}\label{app:NoteOnZeta-NGLs}

As a byproduct, we notice from eqs. \eqref{eq:Aqqbarint} and \eqref{eq:g3} that one may define the following possibly ``new'' logarithmic-integral representation for the Riemann-Zeta function:
\begin{align}\label{eq:Zeta_newFormula}
\zeta_s \equiv \zeta(s) = \frac{(-1)^{s-1}}{\Gamma(s)} \int_0^{+\infty} \ln^{s-1}\left(1 - e^{-\eta} \right) \d\eta\,,
\end{align}
where $\Gamma(s) = (s-1)!$ is the Gamma function and the variable $s$ is greater than 1. The above formula is valid if $s$ is an integer. In the case of non-integer real values one has to take the modulus of the right-hand-side in eq. \eqref{eq:Zeta_newFormula}. In terms of the polar variables $(\theta,\phi)$ the Zeta function admits the integral formula:
\begin{align}\label{eq:Zeta_newFormula_polar}
 \zeta_s = \frac{1}{\Gamma(s)} \int_{-1}^0 \frac{2}{1 - c^2} \ln^{s-1}\left( \frac{1 - c}{-2\,c}\right) \d c\,,
\end{align}
where $c \equiv \cos\theta$. Notice that the form \eqref{eq:Zeta_newFormula_polar} seems to fail (in \texttt{Mathematica 9}) for $s>10$. Moreover, it is interesting to note that the first non-divergent value of the Zeta function is for $s=2$, and so is the first non-vanishing coefficient of NGLs. If we let $\cS_s$ denote the NGLs coefficient at the $s^{\mr{th}}$ loop order then we can write $\cS_s$ as the Mellin transform of the function $(e^\eta - 1)^{-1}$ \cite{Tagaris:Zeta}. That is:
\begin{align}\label{eq:Sn_MellinTransFormula}
 \cS_s = \zeta_s = \frac{1}{\Gamma(s)} \int_0^{+\infty} \frac{\eta^{s-1}}{e^{\eta} - 1}\,\d\eta\,,
\end{align}
which is at least true for two and three-loops NGLs coefficients.
Recall that the Mellin transform techniques were employed in ref. \cite{Catani:1992ua} to compute the first resummed result for event-shape distributions. Whether there exists a more profound relation between NGLs and the Zeta function (and its related functions such as the polylogarithms) is a subject that requires further investigations.
%------------------------------------
\bibliographystyle{JHEP}
\bibliography{refs}
%------------------------------------
\end{document}